 \documentclass{aastex6}

\fullcollaborationName{The Friends of AASTeX Collaboration}

\shorttitle{Asphericity in the Circumstellar Medium of SNe IIn}
\shortauthors{Katsuda et al.}

\begin{document}


\title{Two Distinct-Absorption X-Ray Components from Type IIn Supernovae:\\ 
Evidence for Asphericity in the Circumstellar Medium}


\author{Satoru Katsuda\altaffilmark{1}, Keiichi Maeda\altaffilmark{2,3}, Aya Bamba\altaffilmark{4,5}, Yukikatsu Terada\altaffilmark{6}, Yasushi Fukazawa\altaffilmark{7,8,9}, Koji Kawabata\altaffilmark{7,8}, Masanori Ohno\altaffilmark{7,9}, Yasuharu Sugawara\altaffilmark{10}, Yohko Tsuboi\altaffilmark{1}, and Stefan Immler\altaffilmark{11,12}
}

\altaffiltext{1}{Department of Physics, Faculty of Science \& Engineering, Chuo University, 1-13-27 Kasuga, Bunkyo, Tokyo 112-8551, Japan; katsuda@phys.chuo-u.ac.jp}

\altaffiltext{2}{Department of Astronomy, Kyoto University, Kitashirakawa-Oiwake-cho, Sakyo-ku, Kyoto 606-8502, Japan}

\altaffiltext{3}{Kavli Institute for the Physics and Mathematics of the Universe (WPI), University of Tokyo, 5-1-5 Kashiwanoha, Kashiwa, Chiba 277-8583, Japan}

\altaffiltext{4}{Department of Physics, The University of Tokyo, 7-3-1 Hongo, Bunkyo-ku, Tokyo 113-0033, Japan}

\altaffiltext{5}{Research Center for the Early Universe, School of Science, The University of Tokyo, 7-3-1 Hongo, Bunkyo-ku, Tokyo 113-0033, Japan}

\altaffiltext{6}{Graduate School of Science and Engineering, Saitama University, 255 Shimo-Ohkubo, Sakura, Saitama 338-8570, Japan}

\altaffiltext{7}{Department of Physical Science, Hiroshima University, 1-3-1 Kagamiyama, Higashi-Hiroshima, Hiroshima 739-8526, Japan}

\altaffiltext{8}{Hiroshima Astrophysical Science Center, Hiroshima University, 1-3-1 Kagamiyama, Higashi-Hiroshima, Hiroshima 739-8526, Japan}

\altaffiltext{9}{Core Research for Energetic universe (Core-U), Hiroshima University, 1-3-1 Kagamiyama, Higashi-Hiroshima, Hiroshima 739-8526, Japan}

\altaffiltext{10}{Institute of Space and Astronautical Science, Japan Aerospace eXploration Agency, 3-1-1 Yoshinodai, Sagamihara, Kanagawa 252-5210, Japan}

\altaffiltext{11}{Astrophysics Science Division, NASA Goddard Space Flight Center, Greenbelt, MD 2077, USA}

\altaffiltext{12}{Department of Astronomy, University of Maryland, College Park, MD 20742, USA}

\begin{abstract}

We present multi-epoch X-ray spectral observations of three Type IIn supernovae (SNe), SN~2005kd, SN~2006jd, and SN~2010jl, acquired with {\it Chandra}, {\it XMM-Newton}, {\it Suzaku}, and {\it Swift}.  Previous extensive X-ray studies of SN~2010jl have revealed that X-ray spectra are dominated by thermal emission, which likely arises from a hot plasma heated by a forward shock propagating into a massive circumstellar medium (CSM).  Interestingly, an additional soft X-ray component was required to reproduce the spectra at a period of $\sim$1--2 yr after the SN explosion.  Although this component is likely associated with the SN, its origin remained an open question.  We find a similar, additional soft X-ray component from the other two SNe IIn as well.  Given this finding, we present a new interpretation for the origin of this component; it is thermal emission from a forward shock essentially identical to the hard X-ray component, but directly reaches us from a void of the dense CSM.  Namely, the hard and soft components are responsible for the heavily- and moderately-absorbed components, respectively.  The co-existence of the two components with distinct absorptions as well as the delayed emergence of the moderately-absorbed X-ray component would be evidence for asphericity of the CSM.  We show that the X-ray spectral evolution can be qualitatively explained by considering a torus-like geometry for the dense CSM.  Based on our X-ray spectral analyses, we estimate the radius of the torus-like CSM to be on the order of $\sim$5$\times10^{16}$\,cm.

\end{abstract}

\keywords{circumstellar matter --- supernovae: general --- supernovae: individual (SN~2005kd, SN~2006jd, SN~2010jl) --- X-rays: general}

\section{Introduction} \label{sec:intro}

Mass loss from massive stars is directly linked to the massive stars' evolution, affecting a star's apparent temperature, and often luminosity and burning lifetime as well, for most of their lives.  Therefore, it has been actively investigated from both observational and theoretical points of view \citep[e.g.,][]{2014ARA&A..52..487S}.  In particular, the mass loss in the final stage toward SN explosions, which determines the type of a resulting supernova (SN) explosion, is one of the main issues in modern stellar astrophysics.  

One way to study the mass loss at the last stage of the massive star's evolution is to observe young SNe.  Mass loss influences the stellar environments, forming the circumstellar medium (CSM) around the progenitor star.  The CSM will be excited by the collision with the SN ejecta, and emits intense radiation in various wavelengths, allowing us to reveal the CSM properties and the mass-loss history of the progenitor.  In fact, the mass-loss histories have been revealed for a number of various types of SNe, based on radio, optical, and X-ray observations \citep[e.g.,][]{2012MNRAS.419.1515D,2013MNRAS.435.1520M,2014ApJ...797....2K}.  

It is also important to measure the geometry of the CSM around a SN, because a direct comparison with a CSM directly imaged around an evolved massive star can help understand the nature of the progenitor (e.g., whether or not $\eta$ Carinae can explode immediately).  However, little is known about the geometry of the CSM, since extragalactic SNe are spatially unresolvable during the early phase of their evolution.  This situation has been gradually changing by recent detailed spectropolarimetry of young SNe, first pointed out by \citet{1982ApJ...263..902S}.  However, these measurements are mostly for SN photospheric geometries (i.e., the SN ejecta), and the CSM geometries still remain uncertain \citep[][for a rewiew]{2008ARA&A..46..433W}.

CSM geometries have been revealed/inferred for only a few SNe including a remarkable example, SN~1987A, for which a complex CSM ring was directly imaged \citep[e.g.,][]{1995ApJ...452..680B}.  Others include SNe~1997eg, 1998S, and 2010jl, all of which are classified as Type IIn --- a rare class of SNe comprising of $\sim$9\% of all core-collapse SNe \citep{2011MNRAS.412.1522S}, characterized by an intense narrow H$\alpha$ line, and thought to have experienced the most drastic mass-loss episodes among all types of SNe.  Polarization spectra of the three SNe IIn are strikingly similar with each other in that continuum emission is wavelength-independently polarized by $\sim$2\% and line emission is depolarized \citep{2000ApJ...536..239L,2001ApJ...550.1030W,2008ApJ...688.1186H,2011A&A...527L...6P}.  This result led \citet{2000ApJ...536..239L} and \citet{2008ApJ...688.1186H} to suggest a dense, disk-like or ring-like CSM surrounding aspherical SN ejecta.  We should however note that the interstellar polarization can change ``valleys" into ``peaks" (or vice versa) in the polarization spectrum, as clearly demonstrated by Figure~10 in \citet{2000ApJ...536..239L}.  Since it is not easy to know a correct interstellar polarization, there remain large uncertainties on the CSM geometry.  Therefore, additional observational information about the CSM geometry has been awaited.

We here present new evidence for asphericity in the CSM for three Type IIn SNe, SN~2005kd, SN~2006jd, and SN~2010jl, based on multi-epoch X-ray spectral observations.  Essentially, in the very early-phase, we can detect a heavily-absorbed X-ray component solely, but later ($\sim$1--2\,yr after explosions), an additional moderately-absorbed X-ray component emerges.  The co-existence of the two components and the delayed emergence of the moderately-absorbed component argues for asphericity, presumably a torus-like geometry, in the CSM.  In Section~\ref{sec:obs_ana}, we present information about observations as well as our analyses.  We give our interpretation and conclusion in Sections~\ref{sec:discussion} and \ref{sec:conclusion}, respectively.


\section{Observations and Results} \label{sec:obs_ana}

We analyzed a number of X-ray observations listed in Table~\ref{tab:obs}.  Parts of these observations were already published in the literature \citep{2007ATel..981....1I,2007ATel.1023....1P,2016arXiv160706104D,2012ApJ...755..110C,2012ApJ...750L...2C,2015ApJ...810...32C}, while the rest are presented here for the first time.  As for the {\it Swift} data, we combined several observations taken closely in time to improve the photon statistics.  The explosion dates are taken from the literature: 2005-11-10 for SN~2005kd \citep{2008PZ.....28....6T}, 2006-10-06 for SN~2006jd \citep{2006CBET..679....1B}, and 2010-10-01 for SN~2010jl \citep{2015ApJ...810...32C}.  We utilize the HEAsoft\footnote{http://heasarc.gsfc.nasa.gov/docs/software/lheasoft/} (version 6.16), CIAO\footnote{http://asc.harvard.edu/ciao/} (version 4.6), and SAS\footnote{http://www.cosmos.esa.int/web/xmm-newton/sas} (version 2014-11-04) software packages to analyze the data.  

We extracted a source spectrum from a circular region with a radius of 90$^{\prime\prime}$, 30$^{\prime\prime}$, or 1\farcs5 for each {\it Suzaku}, {\it Swift}/{\it XMM-Newton}, or {\it Chandra} observation, respectively, and subtract background emission from its surrounding annular region.  Within a few 10 arcsec of SN~2010jl, there are faint X-ray point sources including the closest source UGC~5189A \citep[thought to be a galaxy:][]{2015ApJ...810...32C} and six other sources \citep{2015ApJ...810...32C}.  As for {\it Chandra}, we choose the source region to avoid all of these contaminating sources.  On the other hand, spatial resolutions of the other telescopes (i.e., {\it Suzaku}, {\it XMM-Newton}, and {\it Swift}) are not capable of resolving these sources, so that we take account of the contaminating emission in our spectral modeling.  To this end, we adopt two absorbed power-law components for UGC~5189A and the other six sources, following \citet{2015ApJ...810...32C}.  We fix the hydrogen column densities and power-law indices to $N_{\rm H}$ = 4$\times10^{21}$\,cm$^{-2}$ (UGC~5187A) and 5$\times10^{21}$\,cm$^{-2}$ (six nearby sources) and a power-law index of $\Gamma$ = 1.15 (UGC~5187A) or 2.05 (six nearby sources).  The power-law normalizations are set to the {\it Chandra}'s best-fit values closest in time listed in Tables~2 and 3 in \citet{2015ApJ...810...32C}.  Furthermore, we added a gaussian component at $\sim$6.41\,keV (after correcting for the redshift of the host galaxy) for some very early-phase spectra.  This line has been identified as Fe K$\alpha$ arising from the neutral or low ionized iron ions in the CSM \citep{2012ApJ...750L...2C}.  

\begin{figure}
\gridline{\fig{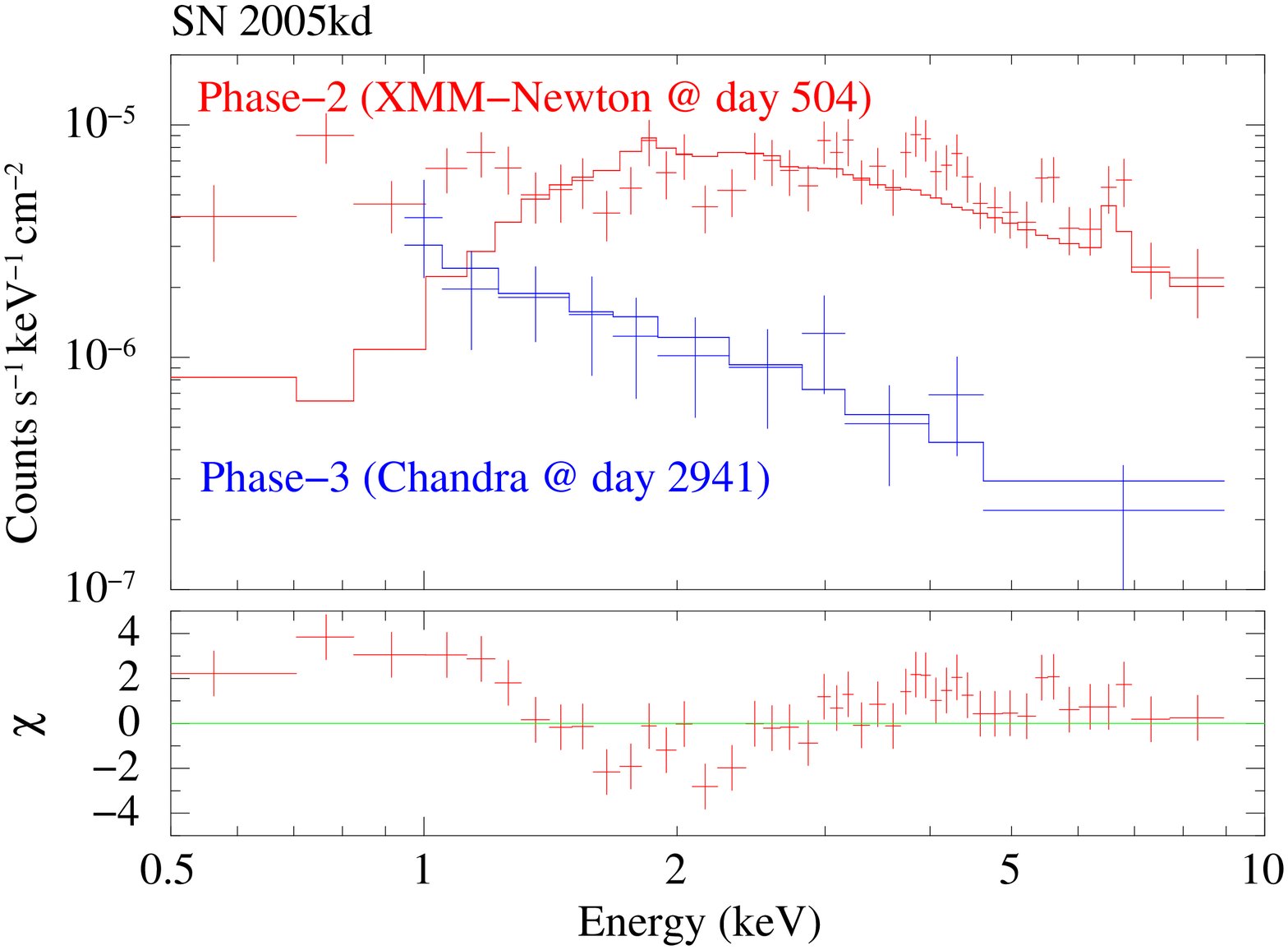}{0.5\textwidth}{}
          \fig{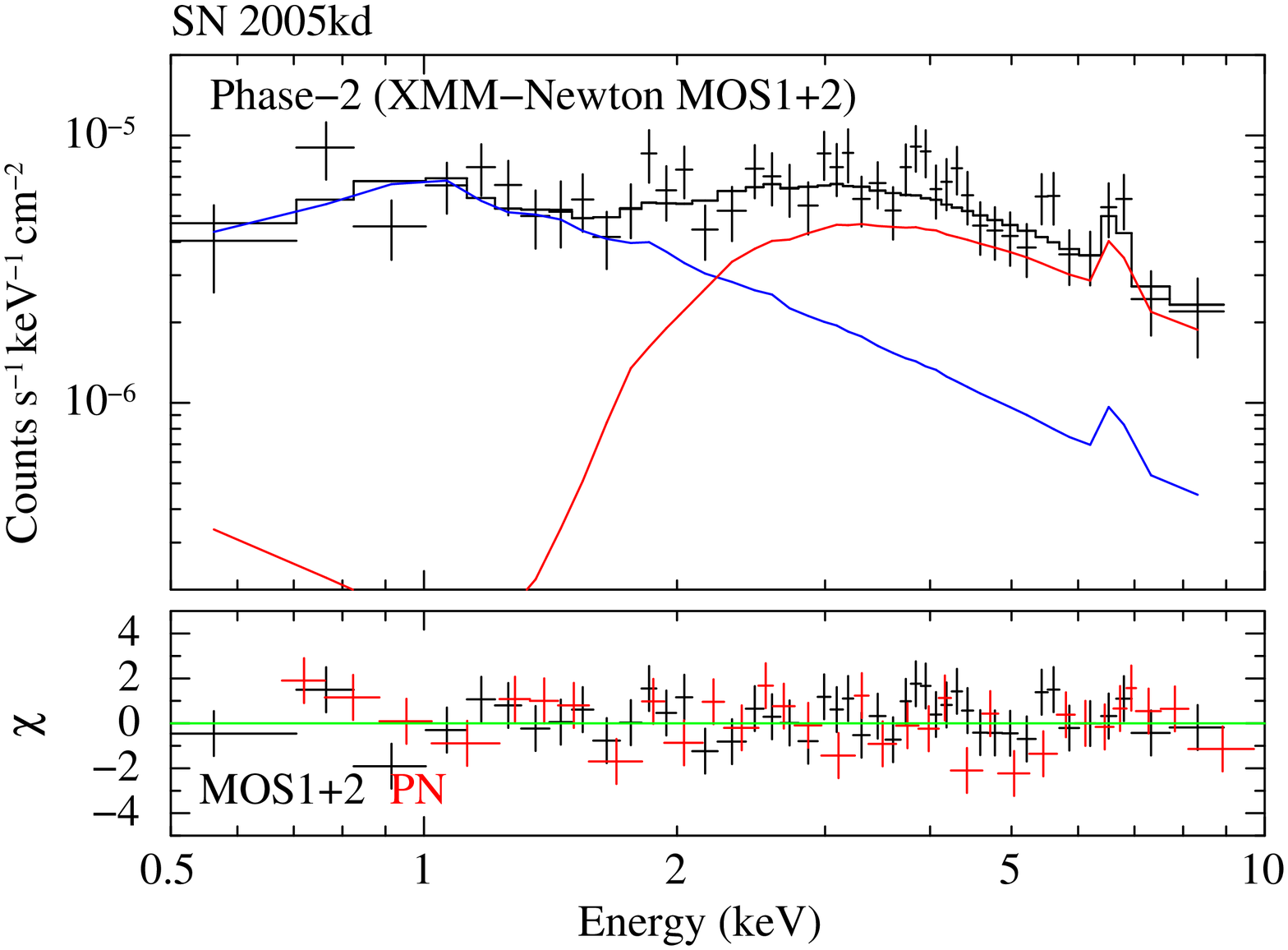}{0.5\textwidth}{}
          }
\gridline{\fig{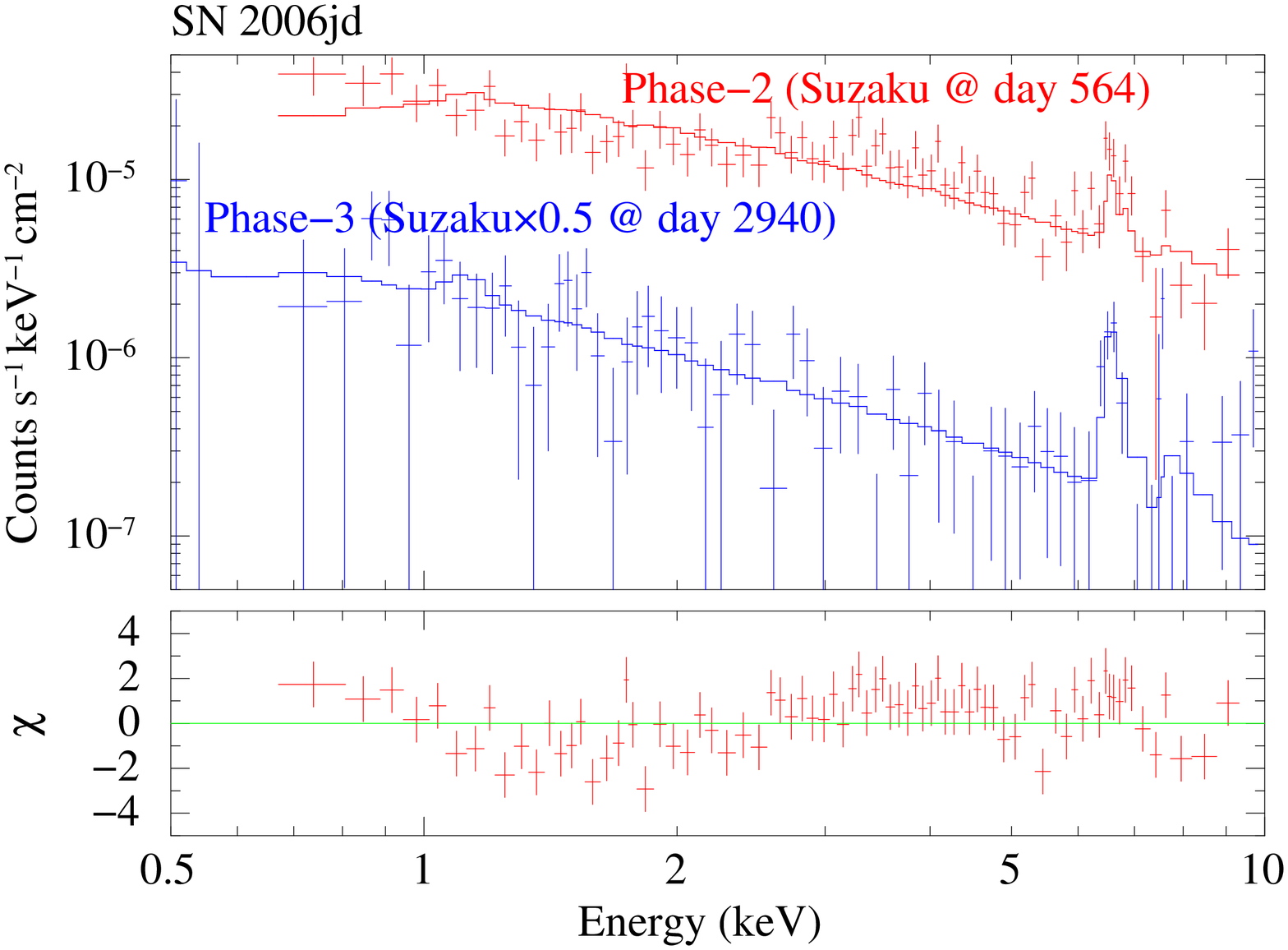}{0.5\textwidth}{}
          \fig{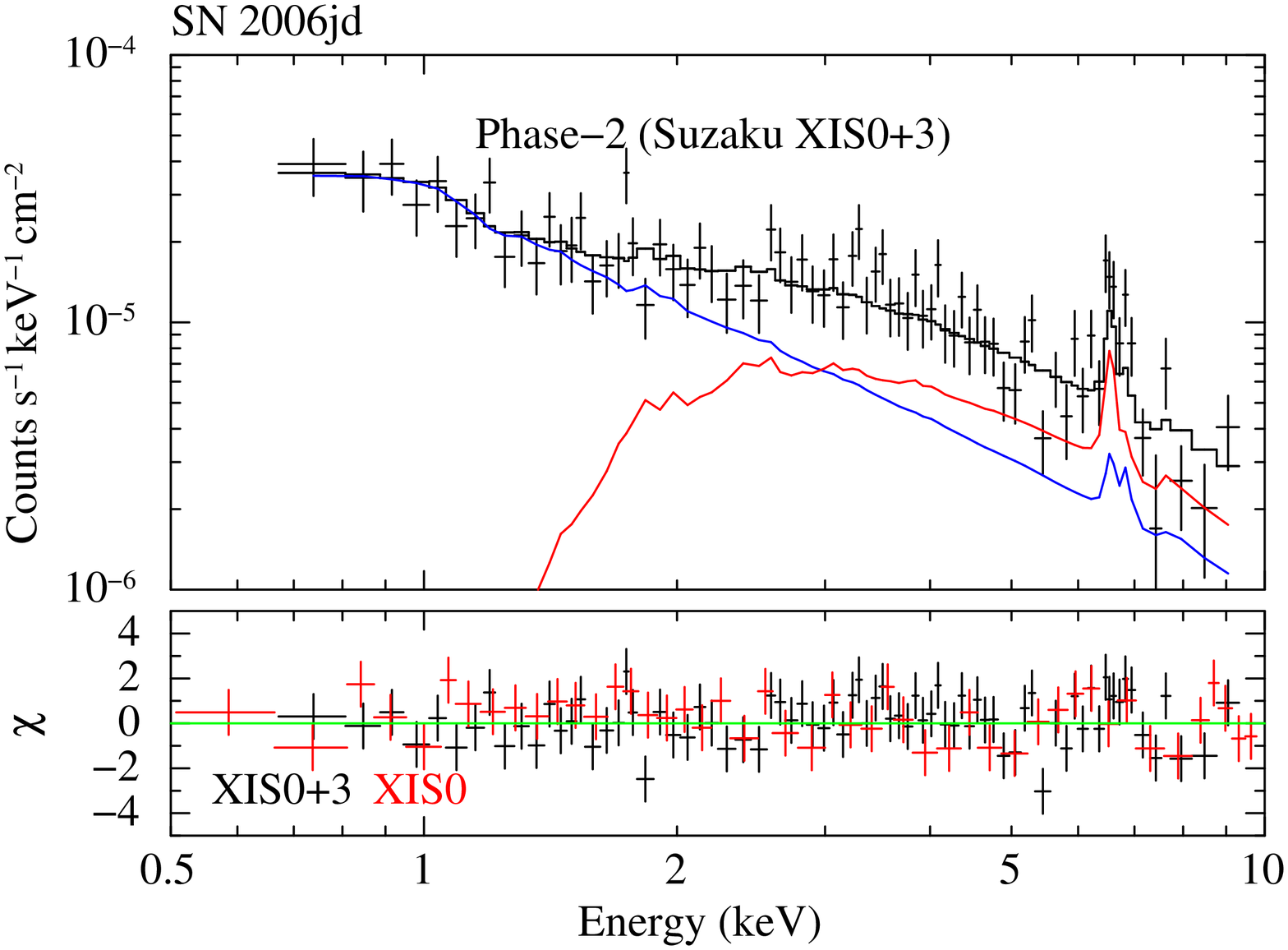}{0.5\textwidth}{}
          }
\gridline{\fig{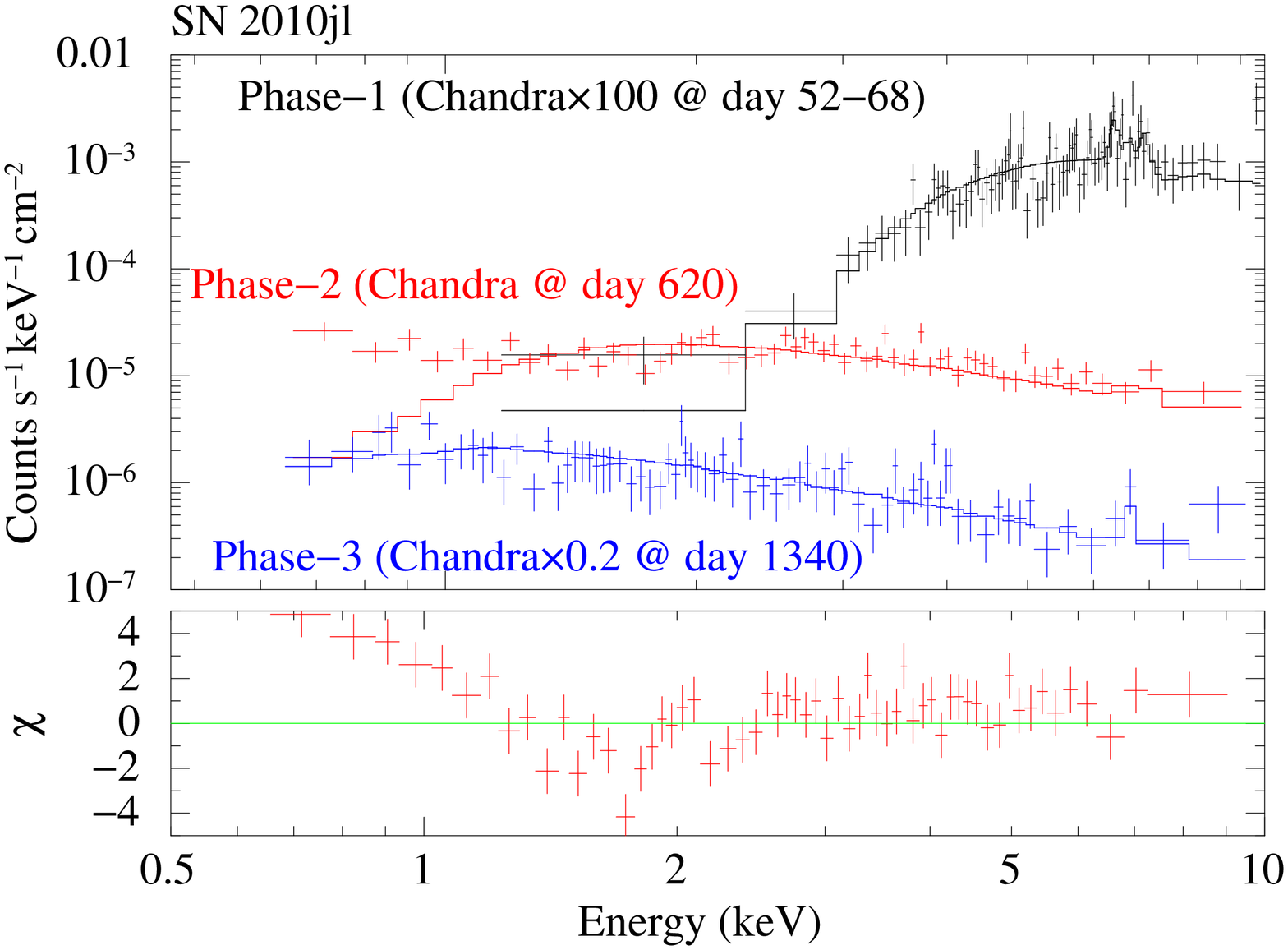}{0.5\textwidth}{}
          \fig{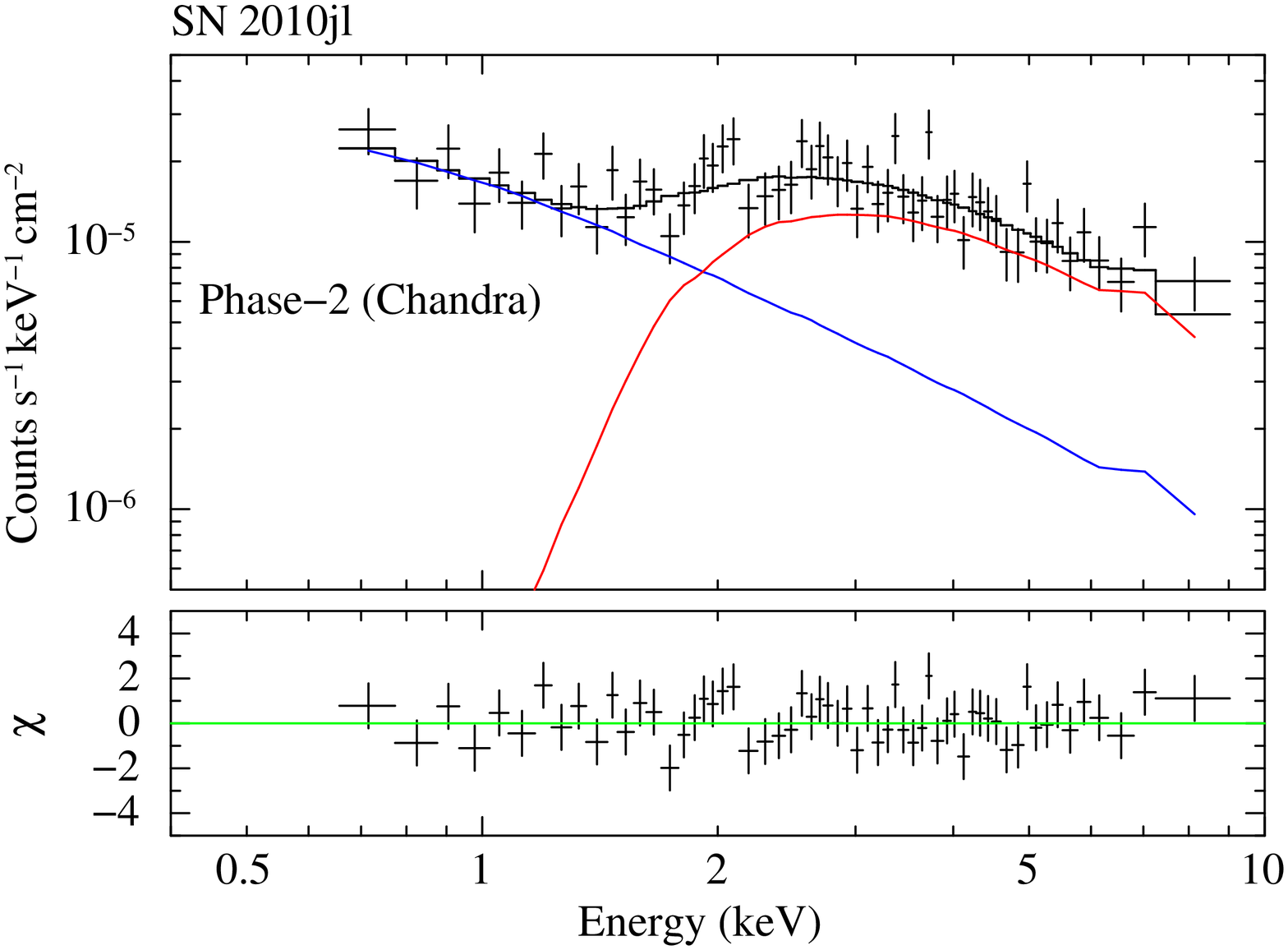}{0.5\textwidth}{}
          }
\caption{Upper left: Typical X-ray spectra in the phases 2 (in red) and 3 (in blue) for SN~2005kd (see Fig.~\ref{fig:csm_image}).  The observed time periods of phases 2 and 3 are days 440--504 and 784--2941, respectively.  The spectra are fitted by a single {\tt vpshock} component model, which results in a good fit for the phase 3, but not for the phase 2, as can be seen in the lower panel showing residuals for the phase 2.  Upper right: The phase-2 spectrum fitted by a two {\tt vpshock} component model.  The lower panel shows residuals.  Note that we do not show the PN data in the upper panel for clarity.  
Middle panels: Same as above but for SN~2006jd.  The observed time periods of phases 2 and 3 are days 564--914 and 1073--2940, respectively. 
Bottom panels: Same as above but for SN~2010jl.  The observed time periods of phases 1, 2, and 3 are days 52--209, 383--620, and 764--1545, respectively.  In the left panel, we show the initial-phase (phase-1) spectrum in black as well as those from later phases.  The model with this spectrum contains a gaussian at $\sim$6.45\,keV in addition to the {\tt vpshock} model.
\label{fig:spec}}
\end{figure}

Since most of the data are statistically too poor to perform chi-square tests, each spectrum is grouped into bins with minimum counts of 5, and we use maximum likelihood statistics for a Poisson distribution, the so-called c-statistics \citep{1979ApJ...228..939C}, to find the best-fit model.  On the other hand, there are some exceptions that have relatively rich photon statistics, including the example spectra shown in Fig.~\ref{fig:spec}.  For them, we grouped the spectra into bins with minimum counts of 20, and performed chi-square tests.

We found/confirmed that most of the spectra were successfully reproduced by either an absorbed \citep[{\tt TBabs}:][]{2000ApJ...542..914W}, non-equilibrium ionization model, i.e., the {\tt vpshock} model \citep{2001ApJ...548..820B} or collisional equilibrium model, i.e., the {\tt vapec} model \citep{2001ApJ...556L..91S}, in XSPEC version 12.8.2k \citep{1996ASPC..101...17A}.  Regardless of the model, the temperature was obtained to be $kT\gtrsim$10\,keV.  This suggests that the X-ray emission is dominated by a forward shock propagating into a CSM, which is consistent with previous studies \citep{2012ApJ...755..110C,2012ApJ...750L...2C,2015ApJ...810...32C}.  

In the literature, thermal equilibrium models had been favorably used over non-equilibrium models \citep[e.g.,][]{2012ApJ...755..110C,2014ApJ...781...42O,2016arXiv160706104D}.  This is mainly motivated by the high density of the CSM, of the order of $\sim$10$^6$\,cm$^{-3}$ based on both the optical spectroscopy and the high mass-loss rates inferred from optical and X-ray light curves \citep[e.g.,][]{2012ApJ...756..173S}.  Such a high density can raise the ionization timescale up to 10$^{12}$\,cm$^{-3}$\,s (i.e., a typical equilibration time) within a few months --- much shorter than the time after the explosion.  

We noticed, however, that some of the spectra exhibiting strong Fe K complexes can not be adequately reproduced by thermal equilibrium models without invoking an unusually high Fe abundance (e.g., 10 times the solar values; see Tables~\ref{tab:param1}--\ref{tab:param3}), as already noted by \citet{2012ApJ...750L...2C} for SN~2010jl.  This is because Fe ions are almost fully ionized for equilibrium states at a high temperature ($>$10\,keV).  Instead, these spectra can be well reproduced by non-equilibrium models without invoking unusually high elemental (Fe) abundances.  In principle, non-equilibrium (under-ionized) plasmas should be present behind (collisionless) forward shocks at any time, as shocks continuously take fresh, unshocked medium.  In addition, bremsstrahlung radiation behind a forward shock cools some of the equilibrated (high density) plasmas down to non-X-ray emitting temperatures, increasing the relative importance of non-equilibrium plasmas compared with the equilibrium plasmas.  For example, the cooling time due to bremsstrahlung radiation, $t_{\rm cool} \sim 1.7\times10^{11} T_{\rm rev}^{0.5}/n_{\rm e}$\,s \citep{1996ApJ...461..993F}, is indeed as short as $\sim$8\,yr for a temperature of 20\,keV and a density of $10^7$\,cm$^{-3}$ (which is expected at 3$\times$10$^{16}$\,cm for a mass-loss rate of 0.01\,M$_\odot$\,yr$^{-1}$ and a wind speed of 100\,km\,s$^{-1}$).  Although it is difficult to estimate the amount of cooling plasmas, as it depends on the density structure of the CSM which is often poorly constrained (see Section 3), cooling would be at least qualitatively non-negligible for forward shocks of SNe~IIn.  

In this context, it is likely that post-shock regions where X-rays are emitted has a range of ionization timescale.  Therefore, we basically adopt a non-equilibrium ionization ({\tt vpshock}) model for our following spectral analyses.  This model takes account of a range of the ionization parameter, where we take a range from zero up to a fitted maximum value.  However, we note that our main conclusion is essentially independent of the model used (whether the equilibrium model or non-equilibrium models) at the expense of an unusually high Fe abundance for the equilibrium model.

Since the electron temperature was not well constrained owing to both low sensitivities at high energies of the X-ray instruments and the relatively poor photon statistics, we fixed the temperature of the {\tt vpshock} component to be $kT_{\rm e}$ = 20\,keV (unless otherwise stated), which is the only example of the robustly-measured temperature of a forward-shocked gas, based on a broadband X-ray spectrum of SN~2010jl with {\it NuSTAR} and {\it XMM-Newton} \citep{2014ApJ...781...42O,2015ApJ...810...32C}.  We also fixed metal abundances to 0.4 times the solar values \citep{2000ApJ...542..914W}, based on metallicities of the SN environments: SN~2006jd \citep{2012ApJ...756..173S}, SN~2010jl \citep{2011ApJ...730...34S}, and SN~2005kd \citep[][for a general trend of long-lasting SNe IIn]{2015A&A...580A.131T}.  An exception is the final-epoch {\it Suzaku} observation of SN~2006jd when the data required an Fe abundance of 1.5--6 solar (see Table~\ref{tab:all2}), even if we use the {\tt vpshock} model.

We found significant deviations from the single-component {\tt vpshock} (or {\tt vapec}) model for some early-phase spectra, as can be seen in the left panels of Fig.~\ref{fig:spec}.  Therefore, we added another component, for which both power-law and thermal components are equally allowed from a statistical point of view.  Based on the $F$-test, statistical significances of adding the second component exceed 99.9\% for all of the three SNe.  The fit results are summarized in Tables~\ref{tab:param1}--\ref{tab:param3}.  As shown in these tables, we examined two electron temperatures for the additional thermal ({\tt vpshock}) component; $kT_{\rm e}$ = 0.5\,keV and 20\,keV representative of the reverse and forward shocks, respectively, which will be discussed below.  We point out that the main fit results (i.e., absorption column densities and the luminosities) are in good agreements between the {\tt vpshock} and {\tt vapec} models, assuring that our analysis is fitting-model independent.  The right panels of Fig.~\ref{fig:spec} show the best-fit models for the double high-temperature {\tt vpshock} case.  

Now, let us examine the validity of the three possibilities (power-law, low-T thermal, high-T thermal) for the additional component, based on astrophysical points of view.  The power-law component could arise from the inverse Compton (IC) scattering of the SN light ($\sim$1\,eV) by accelerated electrons (Lorentz factor $\gamma$ $\sim$30) and/or synchrotron radiation from accelerated electrons.  The former mechanism was discussed in detail to explain the X-ray emission from Type IIb SN~2011dh \citep{2012ApJ...752...78S,2012ApJ...758...81M}.  For SN~2011dh, the IC process was preferred especially because this model was able to reproduce the observed X-ray light curve evolution strikingly well.  However, there was a difficulty in explaining the X-ray luminosity ($\sim$2$\times10^{39}$\,erg\,s$^{-1}$); the electron number density responsible for the IC emission ($\gamma \sim 30$) is required to be two orders of magnitude larger than the extrapolation from the radio spectrum ($\gamma \sim 50-200$), requiring an additional low-energy electron population \citep{2012ApJ...758...81M}.  Later, \citet{2014ApJ...785...95M} analyzed a deep X-ray observation at $\sim$500\,day, and found that the X-ray spectrum at this time is dominated by the reverse-shock thermal emission.  By extrapolating the flux to the early phase, they found that the reverse-shock thermal emission could substantially contribute to the early-phase X-ray spectrum as well, which makes the required IC flux smaller.  Meanwhile, the X-ray luminosities of the SNe IIn in this paper are 1--2 orders of magnitude higher than that of SN~2011dh, whereas the peak radio fluxes are only a few times higher than (or consistent with) that of SN~2011dh.  Therefore, we conclude that the IC scenario is unlikely.  Likewise, the synchrotron possibility is not likely, either.  A simple extrapolation of the radio emission to the X-ray regime results in $\nu L_\nu$ (X-ray) $\sim$ $\nu L_\nu$ (radio) $\sim$ 10$^{37}$\,erg\,s$^{-1}$, which is four orders of magnitude lower than those observed (see Tables~\ref{tab:param1}--\ref{tab:param3}).  If there is a break in the power-law spectrum as is usually seen in SN remnants \citep[e.g.,][]{1999ApJ...525..368R}, the X-ray flux would become even smaller than the simple extrapolation from the radio band, making the synchrotron case very unlikely.

Second, we turn to the low-T ($kT_{\rm e} = 0.5$\,keV) thermal case.  We suppose that this emission would originate from either a reverse shock or a dense clumpy/shell CSM.  In the reverse-shock case, a reasonable density profile of the ejecta ($\rho \propto r^{-10}$) and a forward-shock speed of $\sim$5,000\,km\,s$^{-1}$ indicates the reverse shock to be radiative \citep{2009A&A...494..179N}.  Then, the unabsorbed X-ray luminosity decreases as $L_{\rm rev} \propto t^{-(15 - 6s + sn - 2n)/(n - 3)}$ \citep{1996ApJ...461..993F}, where $t$, $s$, and $n$ are the time after the explosion, and density slopes of the CSM and the SN ejecta $\rho \propto r^{-s, -n}$, respectively.  By substituting $n = 10$ and $s = 2$, which are typical values for Wolf-Rayet stars \citep{2003LNP...598..171C}, we obtain $L_{\rm rev} \propto t^{-0.43}$.  On the other hand, if we fit the multi-epoch spectra with this two-component thermal model, we find that the low-T component quickly disappears as $L_{\rm X} \propto t^{-4}$ (which is not shown in this paper).  This decline is much faster than expected.  In addition, the fact that the absorbing column density of the low-T component is smaller than that of the high-T component (see Table~\ref{tab:param1}--\ref{tab:param3}) cannot be explained by the reverse-shock scenario, since a cold dense shell forming between the reverse shock and the contact discontinuity absorbs the reverse-shock emission, whereas it cannot absorb the forward-shock emission \citep[e.g.,][]{2004MNRAS.352.1213C}.  Thus, we do not favor this idea.  As for the clumpy/shell CSM scenario, such a density inhomogeneity is generally expected in a high-density CSM rather than a low-density CSM.  This would conflict with the fact that the low-T component has a lower absorption than the high-T component (see Tables~\ref{tab:param1}--\ref{tab:param3}).

\begin{figure}
\gridline{\fig{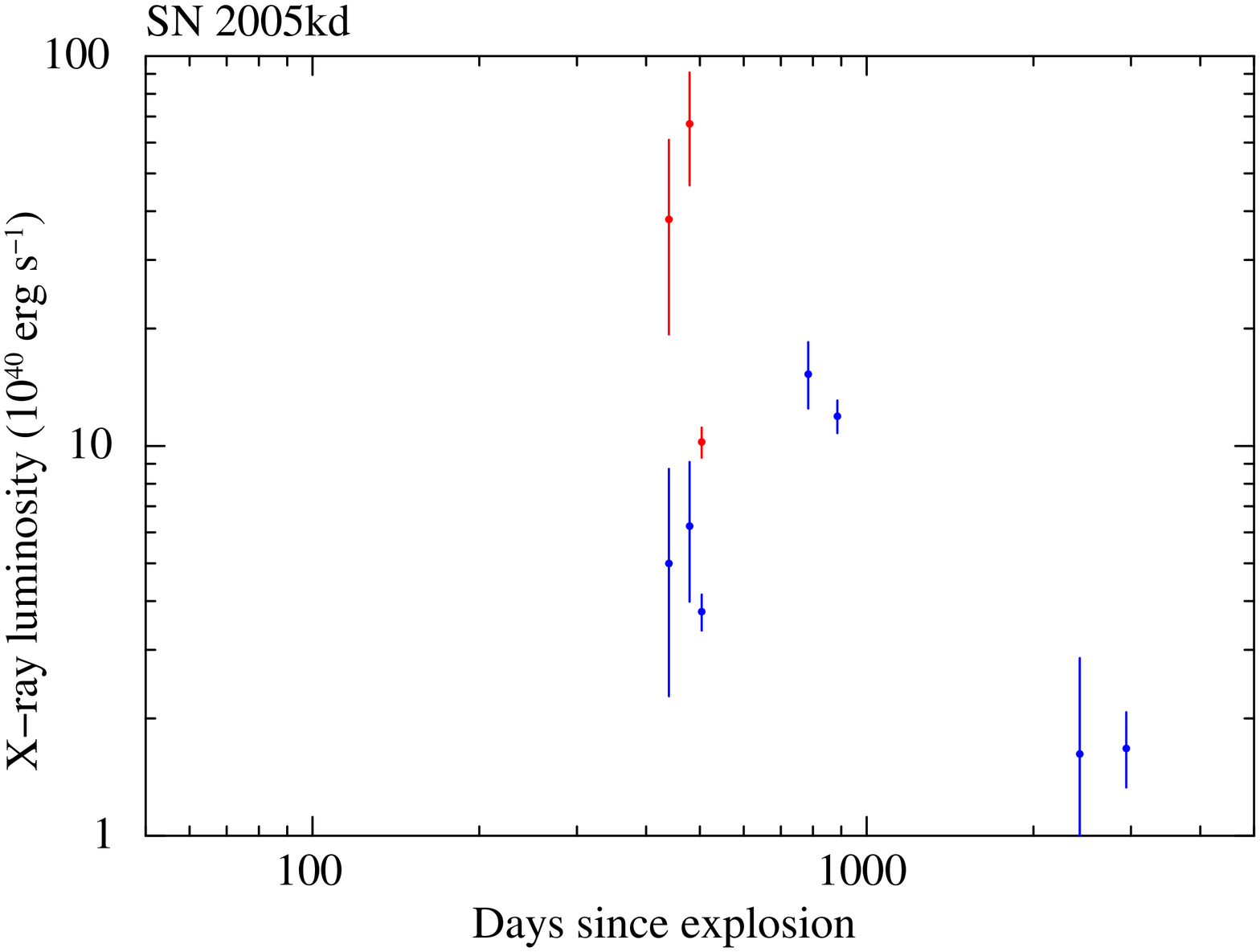}{0.5\textwidth}{}
          \fig{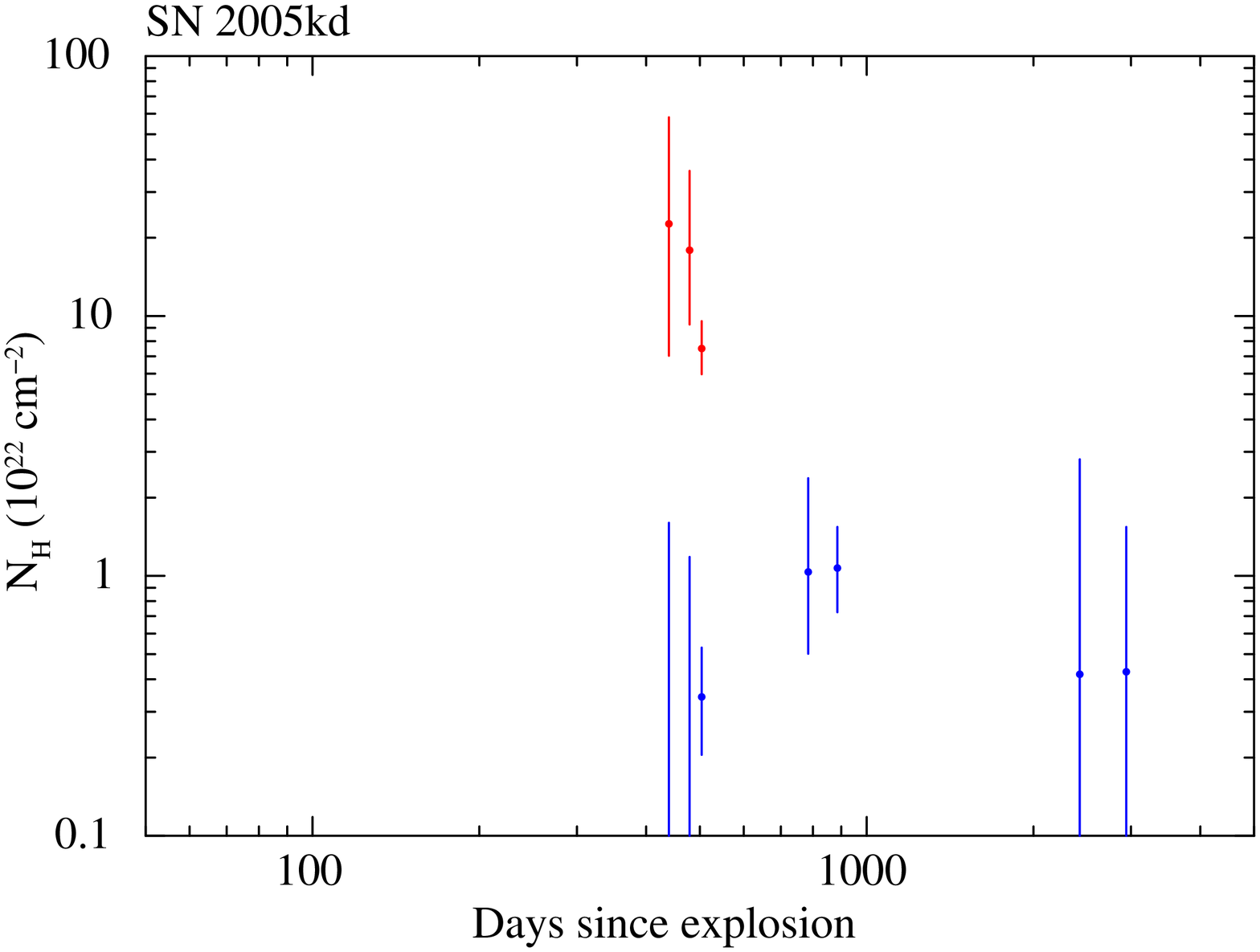}{0.5\textwidth}{}
          }
\gridline{\fig{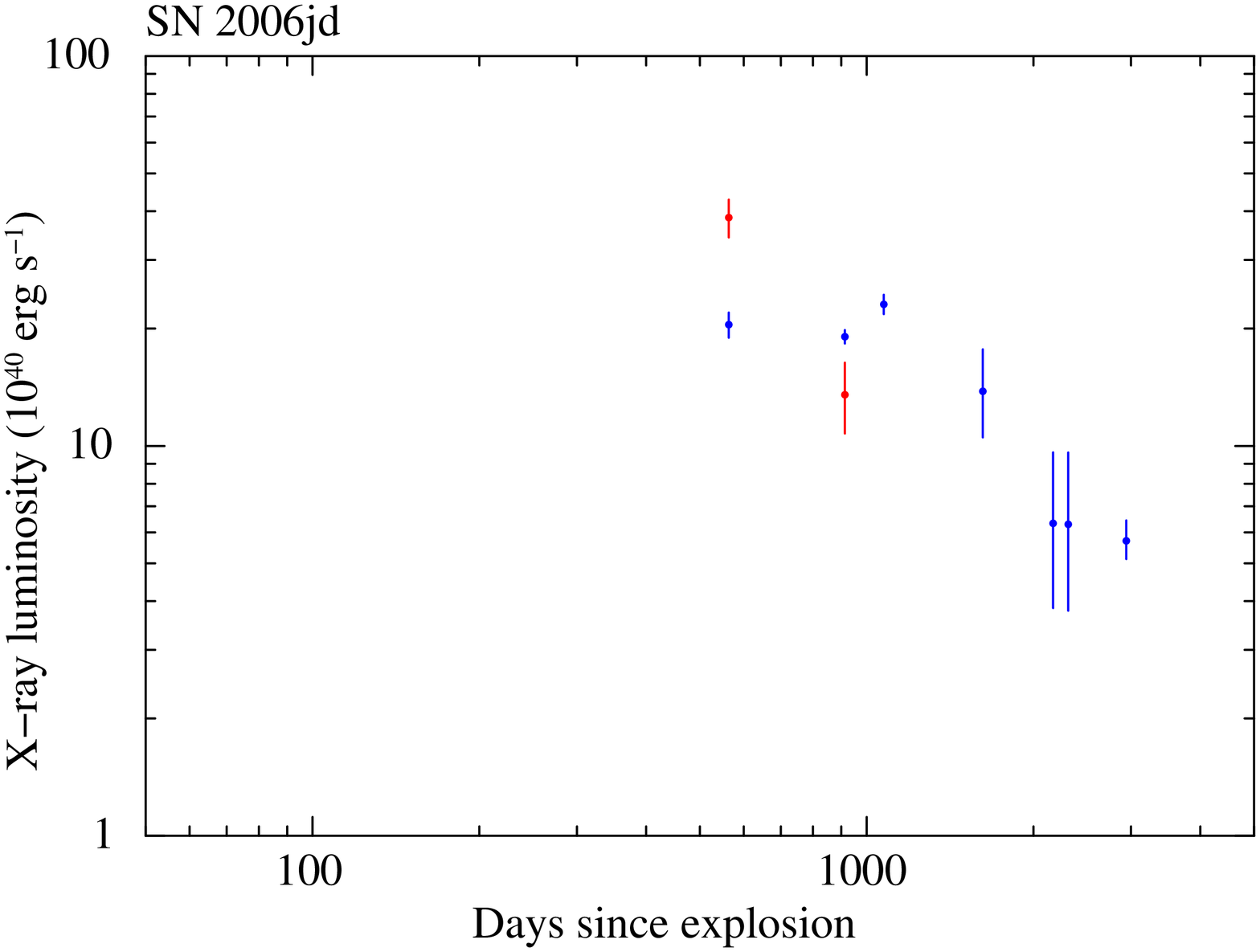}{0.5\textwidth}{}
          \fig{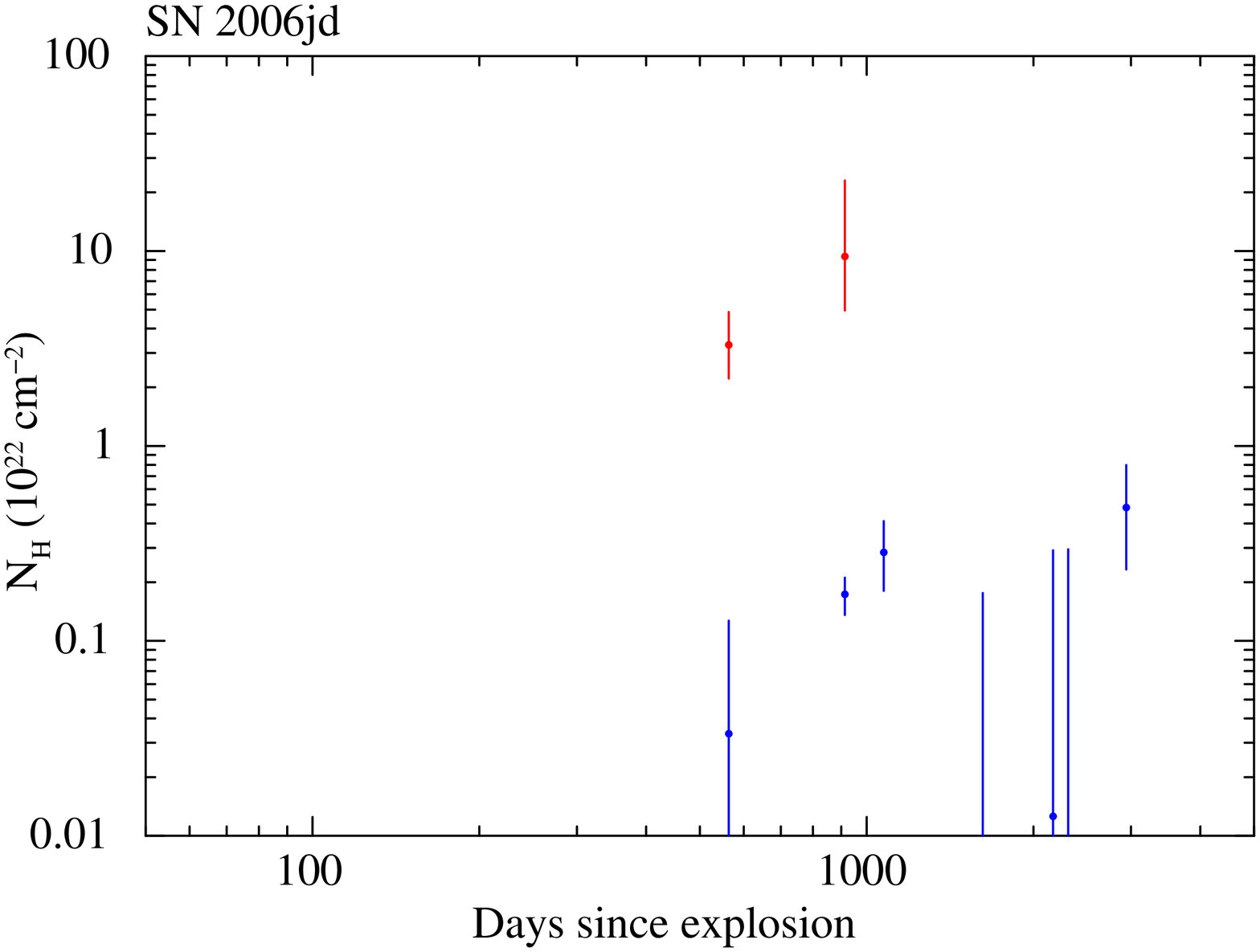}{0.5\textwidth}{}
          }
\gridline{\fig{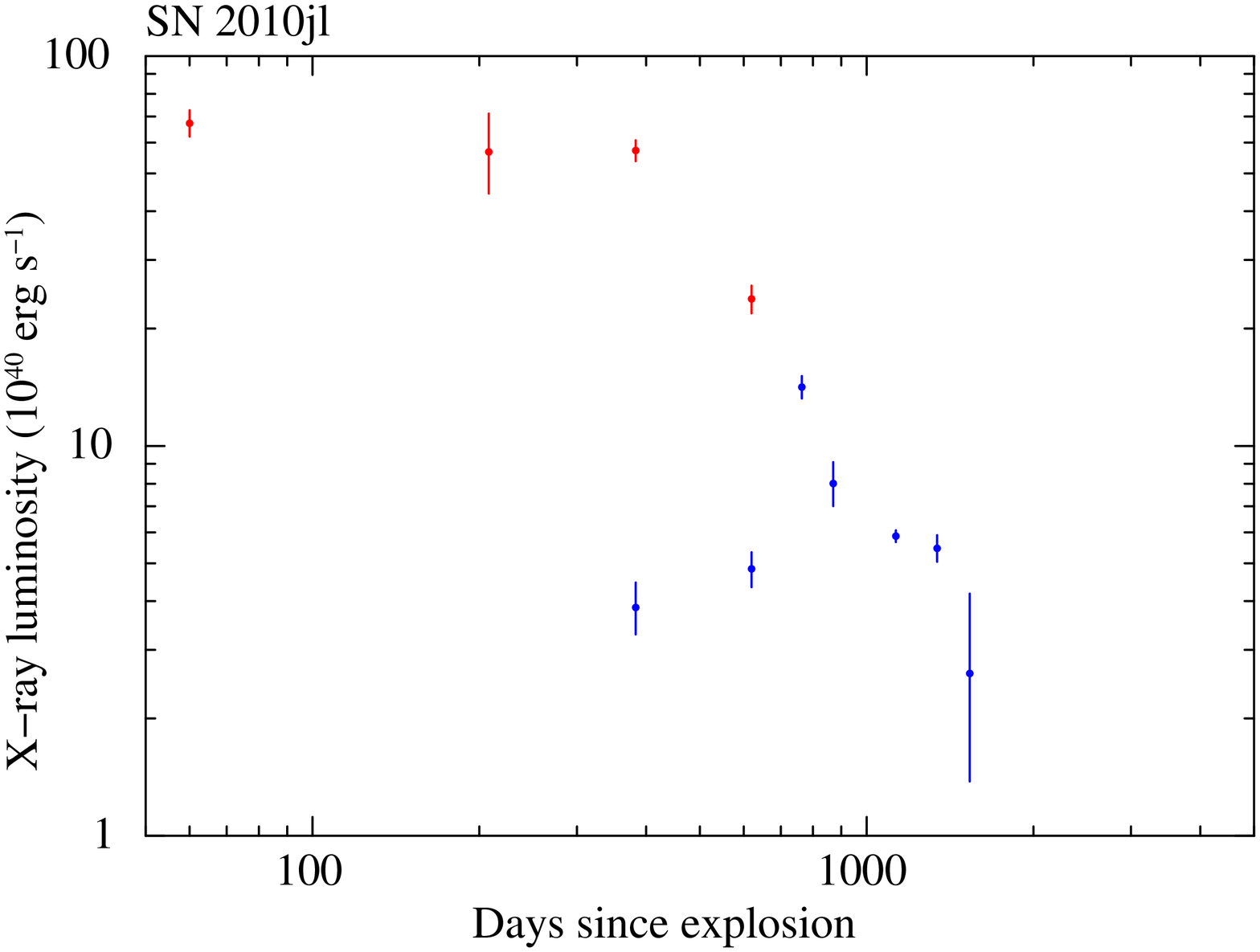}{0.5\textwidth}{}
          \fig{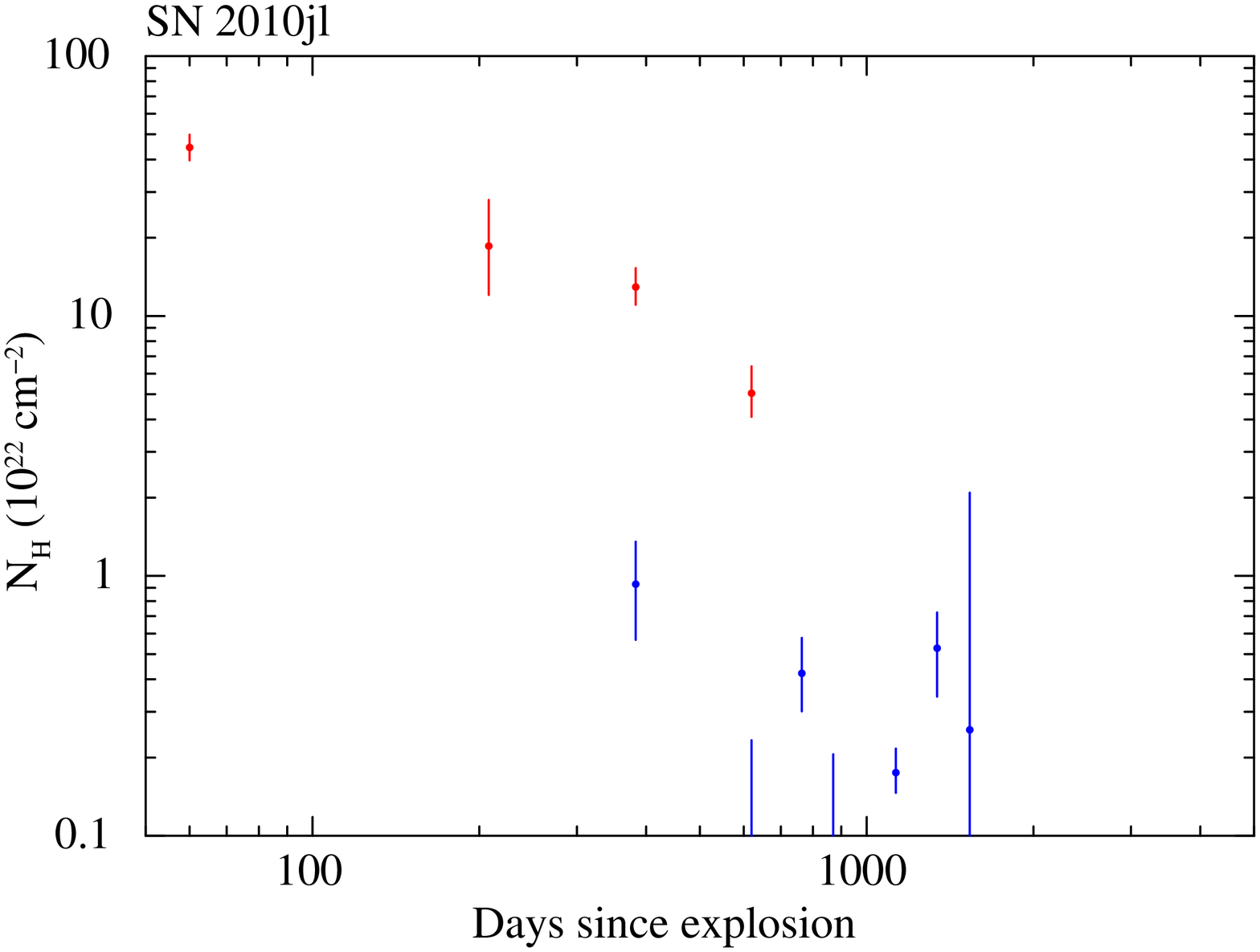}{0.5\textwidth}{}
          }
\caption{Intrinsic (absorption-corrected) X-ray luminosity (left panels) and intervening absorbing material, $N_{\rm H}$, (right panels) as a function of time.  The data in red and blue are responsible for the heavily- and moderately-absorbed thermal {\tt vpshock} component with a high temperature of $kT_{\rm e} = 20$\,keV.  
\label{fig:lumi_nh}}
\end{figure}

In this context, the only viable possibility is the high-temperature ($kT_{\rm e} = 20$\,keV) thermal case, which indeed seems plausible as described below.  Figure~\ref{fig:lumi_nh} shows X-ray light curves and time evolution of absorbing hydrogen column densities for the three SNe, where the data in red and blue are responsible for heavily- and moderately-absorbed thermal ($kT_{\rm e} = 20$\,keV) {\tt vpshock} component.  The fit results responsible for these images are summarized in Tables~\ref{tab:all1}--\ref{tab:all3}.  Note that we fix the ionization timescale to 2$\times$10$^{12}$\,cm$^{-3}$ for very poor statistics data.  We will give our interpretation of this spectral evolution in the next section.

\section{Discussion} \label{sec:discussion}

We have studied X-ray spectral evolution of three SNe IIn (SN~2005kd, SN~2006jd, and SN~2010jl).  Initially, the spectra can be represented by a single, heavily-absorbed thermal component (Phase-1).  Subsequently, the spectra start to show two components, comprised of one heavily absorbed, and another moderately absorbed high-temperature component (Phase-2).  Finally, the heavily absorbed component disappears, leaving only the moderately-absorbed component (Phase-3).  As discussed in the previous section, both of the two components should be associated with the forward shock propagating into the CSM.  The heavily-absorbed component comes through the massive CSM until the forward shock overtakes the dense CSM region, whereas the moderately-absorbed component directly reaches us.

The timing when the moderately-absorbed component emerges is critical to determine the geometry of the massive CSM.  For a spherically symmetric CSM, we expect that a heavily absorbed component, which is usually responsible for the reverse-shock emission, emerges later than the less absorbed component, which is usually responsible for the forward-shock emission.  This is in stark contrast to the spectral evolution observed for the three SNe IIn.  The unexpected X-ray spectral evolution observed, i.e., a heavily absorbed component appears first, and then a less-absorbed component emerges later, requires an aspheircal CSM and/or a density inhomogeneity of the massive CSM (e.g., a spherical CSM having lots of voids/holes), providing us with a new observational support to the mounting evidence for the asphericity of the CSM.  

\begin{figure}[ht!]
\gridline{\fig{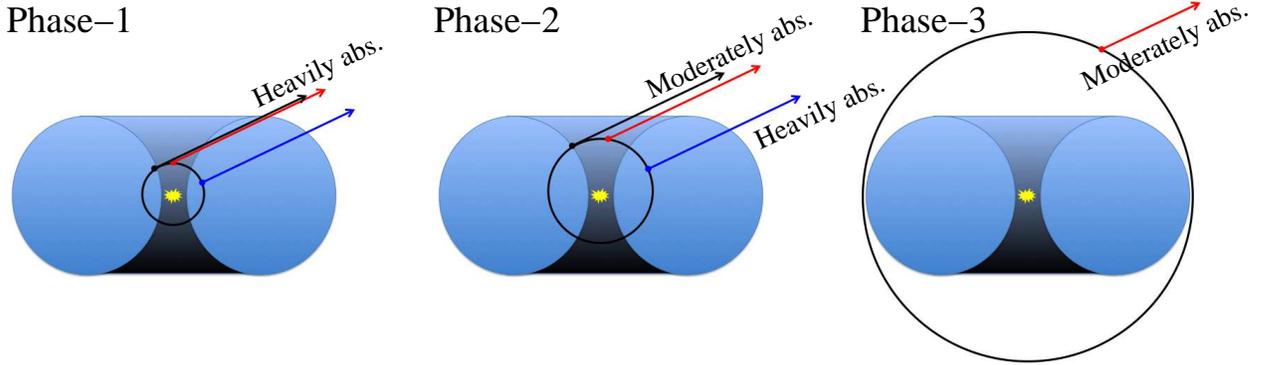}{1.0\textwidth}{}
          }
\caption{Schematic view of the dense CSM torus shown in cross section.  The forward shock is illustrated as a black solid line.  In the initial phase (Phase-1), the forward-shock emission is heavily absorbed in any paths.  In the next phase (Phase-2), part of the forward shock emission can be directly seen.  In the last phase (Phase-3), the forward shock exits the dense CSM region, so that heavily-absorbed component disappears.
\label{fig:csm_image}}
\end{figure}

Specific geometries of the CSM around SNe IIn have been suggested to be disk-like or ring-like, based on previous spectroscopic and spectropolarimetric observations of SNe IIn \citep{2000ApJ...536..239L,2008ApJ...688.1186H}.  We here show that the X-ray spectral evolution can be reasonably explained along the same line.  Figure~\ref{fig:csm_image} illustrates a schematic view of the CSM geometry as well as the origin of the X-ray emission and its path to the observer.  Arrows in black and blue are, respectively, responsible for emission from the far and near sides of the torus hit by the forward shock, while arrows in red are responsible for the emission from the forward shock propagating into the void of the torus.  

In the initial phase (Phase-1), the radius of the forward shock is so small that the entire emission is heavily absorbed by the near-side of the CSM torus.  This phase is expected only if the CSM torus is large and thick enough.  In fact, at least for SN~2010jl, a large covering fraction of the CSM is suggested by the large equivalent width (0.2$\pm$0.1\,keV) of the fluorescent Fe K$\alpha$ \citep{2012ApJ...750L...2C}; EW$_{\rm Fe}$ $\sim$ 0.12 $\left(\frac{f_{\rm c}}{1}\right)$ $\left(\frac{N_{\rm H}}{4\times10^{23} {\rm cm}^{-2}}\right)$ keV, where $f_{\rm c}$ is a covering fraction \citep[e.g.,][]{2007ApJ...665..209M}.  When the forward-shock radius increases enough to be observed over the CSM torus (Phase-2), we can see both of the moderately-absorbed and heavily-absorbed X-ray components.  Here, the moderately-absorbed X-ray component could arise from the hot gas in the void as well as the far-side torus hit by the forward shock.  In the last stage (Phase-3), the forward shock breaks out the dense CSM torus, leaving only the moderately-absorbed X-ray component.  

Given that the X-ray intensity scales as the square of the density, our interpretation that the moderately- and heavily-absorbed components originate from the ``tenuous void" and the ``dense torus" regions, respectively, is in reasonable agreement with the fact that the latter component is brighter than the former component (see Tables~\ref{tab:param1}--\ref{tab:param3}).  Also in this interpretation, we expect a somewhat higher temperature for the moderately-absorbed component than for the heavily-absorbed one due to the possible difference in the shock speed.  However, we assume a temperature of 20\,keV, which is a measurement by {\it NuSTAR}, for both components.  This is because such a high temperature is not sensitive to our X-ray data covering an energy range of 0.5--10\,keV.  Future wide-band spectroscopy may be able to resolve the temperature between the two components.

The relative intensity between the moderately- and heavily-absorbed X-ray components also depends on the viewing angle of the CSM torus.  If viewed edge-on, the moderately-absorbed X-ray component will be very weak, which may be the case for another Type IIn SN~2005ip that do not show the two-component phase \citep{2014ApJ...780..184K}.  If viewed face-on, the moderately-absorbed X-ray component can be seen even earlier than the heavily-absorbed X-ray component.  In this way, future systematic X-ray spectral observations of SNe IIn will help constrain the CSM geometry.  Also, synergies between X-ray observations and optical spectropolarimetry/spectroscopy are very important.

We now constrain properties of the dense CSM torus, based on the results from X-ray spectral analyses.  The timing when the heavily-absorbed component disappeared allows us to estimate the size of the CSM torus.  Assuming that the forward-shock speed is constant at 5,000--10,000\,km\,s$^{-1}$ \citep[][]{2012ApJ...756..173S,2014ApJ...781...42O,2015ApJ...810...32C}, we estimate the radius of the CSM torus to be 3--6$\times$10$^{16}$\,cm or a few 1,000 AU.  

The mass-loss rate for the eruption that produced the CSM torus can be also inferred by comparing the X-ray light curve with an analytical model \citep[Equation 3.18, integrated from 0.2 keV to 10 keV in][]{1996ApJ...461..993F}.  This analytical model assumes a spherically symmetric CSM, which is not likely the case at least for the three SNe IIn.  However, if the covering fraction of the CSM is large enough (which is indicated by the large EW of the Fe K line), the spherically symmetric model is not a bad approximation.  
From the early-phase ($t < 500$\,day) heavily-absorbed X-ray component of SN~2010jl, we derive the mass-loss rate and the density slope in the CSM to be $\dot M$ $\sim$ 0.01\,($V_w/100$\,km\,s$^{-1}$)\,M$_\odot$\,yr$^{-1}$ (at $r = 10^{15}$\,cm) and $s = 1.55$, which are consistent with previous estimates based on X-ray observations \citep{2012ApJ...750L...2C}.  These two parameters can be also inferred by modeling the $N_{\rm H}$ evolution \citet{1996ApJ...461..993F}, from which we find a consistent density slope but a smaller mass-loss rate by a factor of 5.  This discrepancy may be explained by a substantial amount of the CSM being fully ionized.  Adopting the flux-based mass-loss rate, we calculate the total mass in the CSM torus to be 3.5--9\,M$_\odot$ for a shock speed of 5,000--10,000\,km\,s$^{-1}$.  Assuming the same density slope ($s = 1.55$), we calculated the masses of the CSM torii for the other two SNe IIn to be 2--6\,M$_\odot$.  

These properties of the CSM torus are quite different from those of SN~1987A and its older cousin SN~1978K \citep{2016MNRAS.458.2063K}; the size and mass are an order of magnitude smaller and two orders of magnitude larger than those of the inner CSM ring of SN~1987A, respectively \citep{1995ApJ...452..680B,2007AIPC..937....3M}.  This implies that the nature of the progenitor stars of SNe IIn is very different from that of SN~1987A and presumably general Type IIP/L SNe.  For SNe IIn, the large amounts of mass eruptions during a short period prior to explosions, $t = 200\left(\frac{t_{\rm br}}{2\,{\rm yr}}\right)\left(\frac{V_{\rm sh}}{10,000\,{\rm km}\,{\rm s}^{-1}}\right)\,\left(\frac{V_{\rm wind}}{100\,{\rm km}\,{\rm s}^{-1}}\right)^{-1}$\,yr where $t_{\rm br}$ is the time for the forward shock to breakout the massive CSM, favor the idea that the progenitors of these SNe IIn look like luminous blue variable stars \citep{2012ApJ...756..173S,2013ApJ...776....5M,2014ApJ...781...42O}.  

We should note, however, that neither the mass-loss rate nor the density slope for SN~2010jl is consistent with those obtained from the optical light-curve modeling: $\dot M$ $\sim$ 0.04--0.3\,($V_w/100$\,km\,s$^{-1}$)\,M$_\odot$\,yr$^{-1}$ and $s \sim 2$ \citep{2013MNRAS.435.1520M,2013ApJ...776....5M,2014ApJ...781...42O}.  (Note that the masses from X-ray and optical measurements are roughly consistent with each other, since the combination of the large mass-loss rate and the steep density slope cancels out that of the small mass-loss rate and the flat density slope.)  This might be due to the imperfect modeling of our X-ray data, especially because the analytical model we employed is applicable only for a spherically symmetric CSM \citep{1996ApJ...461..993F}, which is not the case.  Therefore, for a better description of the X-ray data, we may need to develop sophisticated theoretical models/simulations, which is left for future work.  

\section{Conclusion} \label{sec:conclusion}

Based on multi-epoch spectral analyses of three SNe IIn, SN~2005kd, SN~2006jd, and SN~2010jl, we found that their X-ray spectra were initially explained by a single heavily-absorbed component, and then became comprised of moderately- and heavily-absorbed components during $\sim$1--2 yr after explosions.  Both of the two components are most likely high-temperature thermal emission associated with the forward shock propagating into the CSM, but have distinct absorptions.  This X-ray spectral evolution requires a departure of a spherical symmetry in the CSM.  Specifically, a torus-like geometry of the CSM would qualitatively explain the X-ray spectral evolution.  We estimated that the radius of the torus-like CSM is on the order of $\sim$5$\times10^{16}$\,cm.

\acknowledgments

This work is supported by Japan Society for the Promotion of Science KAKENHI Grant Numbers 16K17673 (S.K.), 26800100 (K.M.), 16K17667 (Y.S.).  The work by K.M.\ is supported by World Premier International Research Center Initiative (WPI Initiative), MEXT, Japan.  


\floattable
\begin{deluxetable}{ccccccc}
\tablenum{1}
\tabletypesize{\tiny}
\tablecaption{Observations used in this paper\label{tab:obs}}
\tablewidth{0pt}
\tablehead{
\colhead{Object} & \colhead{Date (UT)} & \colhead{Day$^a$} & 
\colhead{Instrument} & \colhead{Exposure (ks)} & \colhead{PI} 
& \colhead{References} 
}
\startdata
SN~2005kd & 2007-01-24 & 440 & {\it Swift}/XRT & 8.9 & D.\ Pooley & 1, 2\\
 & 2007-03-04 & 479 & {\it Chandra}/ACIS-S & 3.0 & D.\ Pooley & 2, 3 \\
 & 2007-03-29$^b$ & 504 & {\it XMM-Newton}/MOS1 & 46.0 & N.\ Schartel & 3 \\
 & 2007-03-29$^b$ & 504 & {\it XMM-Newton}/MOS2 & 46.9 & N.\ Schartel & 3 \\
 & 2007-03-29$^b$ & 504 & {\it XMM-Newton}/PN & 8.8 & N.\ Schartel & 3 \\
 & 2008-01-03 & 784 & {\it Chandra}/ACIS-S & 5.0 & D.\ Pooley & 3 \\
 & 2008-04-12 & 886 & {\it Suzaku}/XIS0 & 51 & S.\ Immler & This work \\
 & 2008-04-12 & 886 & {\it Suzaku}/XIS1 & 51.6 & S.\ Immler & This work \\
 & 2008-04-12 & 886 & {\it Suzaku}/XIS2 & 51.6 & S.\ Immler & This work \\
 & 2012-05-18--2012-07-13 & 2380--2436 & {\it Swift}/XRT & 14.3 & R.\ Margutti & 3 \\
 & 2013-11-29$^b$ & 2941 & {\it Chandra}/ACIS-S & 28.7 & V.\ Dwarkadas & 3 \\
\hline
SN~2006jd & 2008-04-22$^b$ & 564 & {\it Suzaku}/XIS0 & 54.2 & S.\ Immler & This work \\
 & 2008-04-22$^b$ & 564 & {\it Suzaku}/XIS1 & 54.2 & S.\ Immler & This work\\
 & 2008-04-22$^b$ & 564 & {\it Suzaku}/XIS3 & 54.2 & S.\ Immler & This work\\
 & 2009-04-07 & 914 & {\it XMM-Newton}/MOS1 & 38.6 & P.\ Chandra & 4\\
 & 2009-04-07 & 914 & {\it XMM-Newton}/MOS2 & 35.8 & P.\ Chandra & 4\\
 & 2009-04-07 & 914 & {\it XMM-Newton}/PN & 32.1 & P.\ Chandra & 4\\
 & 2009-09-14 & 1073 & {\it Chandra}/ACIS-S & 36.8 & P.\ Chandra & 4\\
 & 2011-03-08--2011-03-12 & 1613--1617 & {\it Swift}/XRT & 12.1 & P.\ Chandra & 4\\
 & 2012-09-14--2012-09-17 & 2169--2172 & {\it Swift}/XRT & 7.2 & P.\ Chandra & This work\\
 & 2013-02-02 & 2310 & {\it Swift}/XRT & 9.9 & P.\ Chandra & This work\\
 & 2014-10-24$^b$ & 2940 & {\it Suzaku}/XIS0 & 144.5 & S.\ Katsuda & This work\\
 & 2014-10-24$^b$ & 2940 & {\it Suzaku}/XIS1 & 151.0 & S.\ Katsuda & This work\\
 & 2014-10-24$^b$ & 2940 & {\it Suzaku}/XIS0 & 151.0 & S.\ Katsuda & This work\\
\hline
SN~2010jl & 2010-11-22--2010-12-08$^b$ & 52--68 & {\it Chandra}/ACIS-S & 49.5 & P.\ Chandra \& D. Pooley & 5 \\
& 2011-04-24--2011-04-28 & 205--209 & {\it Swift}/XRT & 10.1 & S.\ Immler & 5\\
 & 2011-10-17 & 383 & {\it Chandra}/ACIS-S & 40.5 & P.\ Chandra & 5, 6 \\
 & 2012-06-10$^b$ & 620 & {\it Chandra}/ACIS-S & 39.5 & P.\ Chandra & 5, 6 \\
 & 2012-11-01 & 764 & {\it XMM-Newton}/MOS1 & 7.5 & N.\ Schartel & 5 \\
 & 2012-11-01 & 764 & {\it XMM-Newton}/MOS2 & 7.9 & N.\ Schartel & 5 \\
 & 2012-11-01 & 764 & {\it XMM-Newton}/PN & 4.9 & N.\ Schartel & 2 \\
 & 2013-01-21--2013-03-29 & 843--910 & {\it Swift}/XRT & 36.4 & E.\ Ofek & 5 \\
 & 2013-11-01 & 1129 & {\it XMM-Newton}/MOS1 & 47.2 & P.\ Chandra & 5 \\
 & 2013-11-01 & 1129 & {\it XMM-Newton}/MOS2 & 47.0 & P.\ Chandra & 5 \\
 & 2013-11-01 & 1129 & {\it XMM-Newton}/PN & 39.9 & P.\ Chandra & 5 \\
 & 2014-06-01$^b$ & 1340 & {\it Chandra}/ACIS-S & 39.5 & P.\ Chandra & 5 \\
 & 2014-11-30--2014-12-24 & 1521--1545 & {\it Swift}/XRT & 16.9 & E.\ Ofek & 5 \\
\enddata
\tablecomments{$^a$Time after the explosion.  $^b$These data are presented in Fig.~\ref{fig:spec}.  References 1--6 are \citet{2007ATel..981....1I,2016arXiv160706104D,2007ATel.1023....1P,2012ApJ...755..110C,2015ApJ...810...32C,2012ApJ...750L...2C}, respectively.}
\end{deluxetable}

\floattable
\begin{deluxetable}{lcccccccc}
\tablenum{2}
\tabletypesize{\tiny}
\tablecaption{Spectral-fit parameters for the phase-2$^a$ data for SN~2005kd\label{tab:param1}}
\tablewidth{0pt}
\tablehead{
\colhead{Model} & \colhead{$N_{\rm H}$ (10$^{22}$\,cm$^{-2}$)} 
& \colhead{$kT_{\rm e}$ (keV)} 
& \colhead{$\Gamma$} 
& \colhead{Fe (solar)} 
& \colhead{log($n_{\rm e}t$/cm$^{-3}$\,s)} 
& \colhead{$L_{\rm X}$ (10$^{40}$ erg s$^{-1}$)}  
& \colhead{$\chi^2$/d.o.f.}  
}
\startdata
{\tt TBabs}$\times${\tt vapec} & 
1.7$\pm$0.4 & 
$>$49 & 
--- &
27$^{+23}_{-14}$ & 
--- &
17.2$\pm$0.9 &
176.3 / 74 \\
{\tt TBabs}$\times${\tt vpshock} & 
2.3$^{+0.5}_{-0.4}$ & 
43$^{+14}_{-8}$ & 
--- & 
0.4$^b$ & 
11.90$^{+0.42}_{-0.20}$ &
12.3$\pm$0.6 &
185.5 / 74 \\
{\tt TBabs}$\times${\tt powerlaw} + {\tt TBabs}$\times${\tt vpshock} & 
0.53$^{+0.68}_{-0.33}$, 6.7$^{+2.2}_{-1.8}$ & 
20$^b$ & 
$2.5^{+2.9}_{-1.3}$ & 
0.4$^b$ & 
11.89$^{+0.54}_{-0.62}$ &
2.89$\pm$0.32, 24.3$\pm$1.6 &
73.6 / 72 \\
{\tt TBabs}$\times${\tt vapec} + {\tt TBabs}$\times${\tt vapec} & 
1.51$^{+0.51}_{-0.55}$, 5.5$^{+1.1}_{-1.0}$ & 
0.5$^b$, 20$^b$ & 
--- & 
0.6$^{+1.1}_{-0.44}$ & 
--- & 
14.4$\pm$1.8, 23.5$\pm$1.2 & 
84.6 / 76 \\
{\tt TBabs}$\times${\tt vapec} + {\tt TBabs}$\times${\tt vapec} & 
0.24$^{+0.13}_{-0.10}$, 7.3$^{+2.0}_{-1.5}$ & 
20$^b$, 20$^b$ & 
--- & 
3.0$^{+1.8}_{-1.5}$ & 
--- & 
4.8$\pm$0.5, 19.2$\pm$1.8 & 
73.6 / 73 \\
{\tt TBabs}$\times${\tt vpshock} + {\tt TBabs}$\times${\tt vpshock} & 
1.5$\pm$0.3, 5.5$^{+1.0}_{-0.9}$ & 
0.5$^b$, 20$^b$ & 
--- & 
0.4$^b$ & 
12.1$^{+0.9}_{-0.5}$ & 
12.1$\pm$1.5, 23.9$\pm$1.4 & 
78.8 / 76 \\
{\tt TBabs}$\times${\tt vpshock} + {\tt TBabs}$\times${\tt vpshock} & 
0.34$^{+0.19}_{-0.14}$, 7.5$^{+2.1}_{-1.5}$ & 
20$^b$, 20$^b$ & 
--- & 
0.4$^b$ & 
12.2$^{+1.1}_{-0.4}$ & 
3.8$\pm$0.4, 10.2$\pm$0.9 & 
76.8 / 73 \\
\enddata
\tablecomments{$^a$The phase-2 ranges from day 440 (or earlier) to 504 after the explosion for SN~2005kd.  The data given in this table were taken at day 504.  $^b$Fixed values.  X-ray luminosities are calculated in a range of 0.2--10\,keV, after correcting for the absorption.  The abundances in the {\tt vpshock} model are fixed to 0.4 solar values, unless otherwise stated.  Two values separated by comma are responsible for individual components, whereas a single value means that the parameter in each component is tied together.
}
\end{deluxetable}

\floattable
\begin{deluxetable}{lcccccccc}
\tablenum{3}
\tabletypesize{\tiny}
\tablecaption{Same as Table~\ref{tab:param1} but for SN~2006jd\label{tab:param2}}
\tablewidth{0pt}
\tablehead{
\colhead{Model} & \colhead{$N_{\rm H}$ (10$^{22}$\,cm$^{-2}$)} 
& \colhead{$kT_{\rm e}$ (keV)} 
& \colhead{$\Gamma$} 
& \colhead{Fe (solar)} 
& \colhead{log($n_{\rm e}t$/cm$^{-3}$\,s)} 
& \colhead{$L_{\rm X}$ (10$^{40}$ erg $^{-1}$)}  
& \colhead{$\chi^2$/d.o.f.}  
}
\startdata
{\tt TBabs}$\times${\tt vapec} & 
0.2$\pm$0.06 & 
$>$47 & 
--- & 
10$^{+5}_{-4}$ & 
--- &
42.7$\pm$1.7 &
192.0 / 122 \\
{\tt TBabs}$\times${\tt vpshock} & 
0.4$^{+0.1}_{-0.2}$ & 
35$^{+21}_{-5}$ & 
--- & 
0.4$^b$ & 
12.06$^{+0.27}_{-0.35}$ &
44.2$\pm$1.7 &
206.2 / 122 \\
{\tt TBabs}$\times${\tt powerlaw} + {\tt TBabs}$\times${\tt vpshock} & 
0.24$^{+0.64}_{-0.18}$, 2.7$^{+1.1}_{-1.3}$ & 
20$^b$ & 
$2.8^{+4.5}_{-1.0}$ & 
0.4$^b$ & 
11.81$^{+0.32}_{-0.30}$ &
2.89$\pm$0.32, 24.3$\pm$1.6 &
136.5 / 120 \\
{\tt TBabs}$\times${\tt vapec} + {\tt TBabs}$\times${\tt vapec} & 
0.76$^{+0.18}_{-0.21}$, 0.9$^{+0.3}_{-0.2}$ & 
0.5$^b$, 20$^b$ & 
--- & 
2.4$^{+0.8}_{-0.7}$ & 
--- & 
27.3$\pm$4.0, 47.1$\pm$2.0 & 
158.1 / 121 \\
{\tt TBabs}$\times${\tt vapec} + {\tt TBabs}$\times${\tt vapec} & 
$<$0.06, 2.8$^{+1.2}_{-0.8}$ & 
20$^b$, 20$^b$ & 
--- & 
1.9$^{+0.8}_{-0.7}$ & 
--- & 
23.1$\pm$1.7, 30.8$\pm$3.9 & 
73.6 / 73 \\
{\tt TBabs}$\times${\tt vpshock} + {\tt TBabs}$\times${\tt vpshock} & 
0.5$^{+0.2}_{-0.3}$, 1.2$^{+0.4}_{-0.3}$ & 
0.5$^b$, 20$^b$ & 
--- & 
0.4$^b$ & 
12.0$^{+0.3}_{-0.2}$ & 
13.9$\pm$1.9, 52.5$\pm$2.3 & 
169.8 / 121 \\
{\tt TBabs}$\times${\tt vpshock} + {\tt TBabs}$\times${\tt vpshock} & 
$<$0.09, 2.8$^{+1.9}_{-0.9}$ & 
20$^b$, 20$^b$ & 
--- & 
0.4$^b$ & 
12.1$^{+0.3}_{-0.2}$ & 
19.8$\pm$1.5, 36.0$\pm$3.8 & 
140.3 / 121 \\
\enddata
\tablecomments{$^a$The phase-2 ranges from day 564 (or earlier) to 914 after the explosion for SN~2006jd.  The data given in this table were taken at day 564.}
\end{deluxetable}

\floattable
\begin{deluxetable}{lcccccccc}
\tablenum{4}
\tabletypesize{\tiny}
\tablecaption{Same as Table~\ref{tab:param1} but for SN~2010jl\label{tab:param3}}
\tablewidth{0pt}
\tablehead{
\colhead{Model} & \colhead{$N_{\rm H}$ (10$^{22}$\,cm$^{-2}$)} 
& \colhead{$kT_{\rm e}$ (keV)} 
& \colhead{$\Gamma$} 
& \colhead{Fe (solar)} 
& \colhead{log($n_{\rm e}t$/cm$^{-3}$\,s)} 
& \colhead{$L_{\rm X}$ (10$^{40}$ erg $^{-1}$)}  
& \colhead{$\chi^2$/d.o.f.}  
}
\startdata
{\tt TBabs}$\times${\tt vapec} & 
1.4$\pm$0.2 & 
$>$53.6 & 
--- & 
22$^{+24}_{-14}$ & 
--- &
23.3$\pm$1.1 &
149.2 /56 \\
{\tt TBabs}$\times${\tt vpshock} & 
1.4$^{+0.3}_{-0.2}$ & 
$>$62 & 
--- & 
0.4$^b$ & 
$>$12 &
21.4$\pm$1.0 &
175.0 / 57 \\
{\tt TBabs}$\times${\tt powerlaw} + {\tt TBabs}$\times${\tt vpshock} & 
$<$0.53, 4.9$^{+1.2}_{-0.9}$ & 
20$^b$ & 
$0.9^{+2.4}_{-0.4}$ & 
0.4$^b$ & 
$>$12 &
8.0$\pm$0.8, 19.6$\pm$2.2 &
51.8 / 54 \\
{\tt TBabs}$\times${\tt vapec} + {\tt TBabs}$\times${\tt vapec} & 
0.5$^{+0.9}_{-0.4}$, 3.6$^{+0.9}_{-0.6}$ & 
0.5$^b$, 20$^b$ & 
--- & 
$<$0.7 & 
--- & 
3.2$\pm$0.4, 28.5$\pm$1.5 & 
63.8 / 55 \\
{\tt TBabs}$\times${\tt vapec} + {\tt TBabs}$\times${\tt vapec} & 
$<$0.16, 4.9$^{+1.4}_{-0.9}$ & 
20$^b$, 20$^b$ & 
--- & 
$<$4.1 & 
--- & 
5.1$\pm$0.5, 24.9$\pm$2.0 & 
51.7 / 55 \\
{\tt TBabs}$\times${\tt vpshock} + {\tt TBabs}$\times${\tt vpshock} & 
1.2$^{+0.3}_{-0.2}$, 3.9$^{+0.8}_{-0.7}$ & 
0.5$^b$, 20$^b$ & 
--- & 
0.4$^b$ & 
$>$12 & 
9.4$\pm$1.2, 28.0$\pm$1.5 & 
67.5 / 55 \\
{\tt TBabs}$\times${\tt vpshock} + {\tt TBabs}$\times${\tt vpshock} & 
$<$0.18, 5.0$^{+1.4}_{-1.0}$ & 
20$^b$, 20$^b$ & 
--- & 
0.4$^b$ & 
$>$12 & 
4.8$\pm$0.5, 23.8$\pm$2.0 & 
52.8 / 55 \\
\enddata
\tablecomments{$^a$The phase-2 ranges from day 383 (or earlier) to 620 after the explosion for SN~2010jl.  The data given in this table were taken at day 620.}
\end{deluxetable}

\floattable
\begin{deluxetable}{lcccccccc}
\tablenum{5}
\tabletypesize{\tiny}
\tablecaption{Best-fit parameters at various epochs of SN~2005kd\label{tab:all1}}
\tablewidth{0pt}
\tablehead{
\colhead{Day$^a$ (Phase)} 
&\colhead{Model} 
& \colhead{$N_{\rm H}$ (10$^{22}$\,cm$^{-2}$)} 
& \colhead{$kT_{\rm e}$} 
& \colhead{Fe (solar)} 
& \colhead{log($n_{\rm e}t$/cm$^{-3}$\,s)} 
& \colhead{$L_{\rm X}$ (10$^{40}$ erg s$^{-1}$)}  
& \colhead{$\chi^2$ or C-value / d.o.f.}  
}
\startdata
440 (2) & 
{\tt TBabs}$\times${\tt vpshock} + {\tt TBabs}$\times${\tt vpshock} & 
$<$1.6, 22.6$^{+35.5}_{-15.6}$ & 
20$^c$, 20$^c$ & 
0.4$^c$ & 
12.3$^c$ &
5.0$^{+3.7}_{-2.7}$, 38.1$^{+22.8}_{-18.8}$ & 
2.3 / 2 \\
479 (2) & 
{\tt TBabs}$\times${\tt vpshock} + {\tt TBabs}$\times${\tt vpshock} & 
$<$1.2, 17.9$^{+18.2}_{-8.6}$ & 
20$^c$, 20$^c$ & 
0.4$^c$ & 
12.3$^c$ &
6.2$^{+2.9}_{-2.2}$, 67.0$^{+23.7}_{-20.4}$ & 
11.4 / 8 \\
504 (2) & 
{\tt TBabs}$\times${\tt vpshock} + {\tt TBabs}$\times${\tt vpshock} & 
0.34$^{+0.19}_{-0.14}$, 7.5$^{+2.1}_{-1.5}$ & 
20$^c$, 20$^c$ & 
0.4$^c$ & 
12.2$^{+1.1}_{-0.4}$ &
3.8$\pm$0.4, 10.2$\pm$0.9 & 
76.8 / 73 \\
784 (3) &  
{\tt TBabs}$\times${\tt vpshock} & 
1.0$^{+1.3}_{-0.5}$ & 
20$^c$ & 
0.4$^c$ & 
12.3$^c$ &
15$\pm$3 & 
7.3 / 12 \\
886 (3) & 
{\tt TBabs}$\times${\tt vpshock} & 
1.1$^{+0.5}_{-0.3}$ & 
20$^c$ & 
0.4$^c$ & 
11.88$^{+0.46}_{-0.32}$ &
12$\pm$1 & 
287.3 / 304 \\
2380--2436 (3) & 
{\tt TBabs}$\times${\tt vpshock} & 
$<$2.8 & 
20$^c$ & 
0.4$^c$ & 
12.3$^c$ &
1.6$^{+1.2}_{-0.9}$ & 
0.5 / 1 \\
2941 (3) & 
{\tt TBabs}$\times${\tt vpshock} & 
$<$1.5 & 
20$^c$ & 
0.4$^c$ & 
12.3$^c$ &
1.7$\pm$0.4 & 
3.4 / 8 \\
\enddata
\tablecomments{$^a$Time after explosion.  $^c$Fixed values.  X-ray luminosities are calculated in a range of 0.2--10\,keV, after correcting for the absorption.  The abundances in the {\tt vpshock} model are fixed to 0.4 solar values, unless otherwise stated.  Two values separated by comma are responsible for individual components, whereas a single value means that the parameter in each component is tied together.}
\end{deluxetable}

\floattable
\begin{deluxetable}{lcccccccc}
\tablenum{6}
\tabletypesize{\tiny}
\tablecaption{Same as Table~\ref{tab:all1} but for SN~2006jd\label{tab:all2}}
\tablewidth{0pt}
\tablehead{
\colhead{Day$^a$ (Phase)} 
&\colhead{Model} 
& \colhead{$N_{\rm H}$ (10$^{22}$\,cm$^{-2}$)} 
& \colhead{$kT_{\rm e}$ (keV)} 
& \colhead{Fe (solar)} 
& \colhead{log($n_{\rm e}t$/cm$^{-3}$\,s)} 
& \colhead{$L_{\rm X}$ (10$^{40}$ erg s$^{-1}$)}  
& \colhead{$\chi^2$ or C-value / d.o.f.}  
}
\startdata
564 (2) & 
{\tt TBabs}$\times${\tt vpshock} + {\tt TBabs}$\times${\tt vpshock} & 
$<$0.09, 3.3$^{+1.6}_{-1.1}$ & 
20$^c$ & 
0.4$^b$ & 
12.1$^{+0.27}_{-0.35}$ & 
20.5$\pm$1.5, 38.5$\pm$4.3 & 
140.3 / 121 \\
914 (2) & 
{\tt TBabs}$\times${\tt vpshock} + {\tt TBabs}$\times${\tt vpshock} & 
0.17$\pm$0.04, 9.4$^{+13.6}_{-4.4}$ &
20$^c$ & 
0.4$^b$ & 
12.3$^{+0.4}_{-0.2}$ & 
19.1$\pm$0.8, 13.5$\pm$2.8 &
227.9 / 203 \\
1073 (3) & 
{\tt TBabs}$\times${\tt vpshock} & 
0.28$^{+0.13}_{-0.10}$ & 
20$^c$ & 
0.4$^b$ & 
$>$12.0 &
23.1$\pm$1.3 & 
43.7 / 36 \\
1613--1617 (3) & 
{\tt TBabs}$\times${\tt vpshock} & 
$<$0.18 & 
20$^c$ & 
0.4$^b$ & 
12.3$^c$ &
13.8$^{+3.9}_{-3.3}$ & 
4.4 / 7 \\
2169--2172 (3) & 
{\tt TBabs}$\times${\tt vpshock} & 
$<$0.28 & 
20$^c$ & 
0.4$^b$ & 
12.3$^c$ &
6.3$^{+3.3}_{-2.5}$ & 
2.0 / 3 \\
2310 (3) & 
{\tt TBabs}$\times${\tt vpshock} & 
$<$0.3 & 
20$^c$ & 
0.4$^b$ & 
12.3$^c$ &
6.3$^{+3.3}_{-2.5}$ & 
8.7 / 2 \\
2940 (3) & 
{\tt TBabs}$\times${\tt vpshock} & 
0.48$^{+0.31}_{-0.25}$ & 
20$^c$ & 
2.5$^{+3.7}_{-1.1}$ & 
12.3$^c$ &
5.7$^{+0.7}_{-0.6}$ & 
177.3 / 188 \\
\enddata
\tablecomments{}
\end{deluxetable}

\floattable
\begin{deluxetable}{lcccccccc}
\tablenum{7}
\tabletypesize{\tiny}
\tablecaption{Same as Table~\ref{tab:all1} but for SN~2010jl\label{tab:all3}}
\tablewidth{0pt}
\tablehead{
\colhead{Day$^a$ (Phase)} 
&\colhead{Model} 
& \colhead{$N_{\rm H}$ (10$^{22}$\,cm$^{-2}$)} 
& \colhead{$kT_{\rm e}$ (keV)} 
& \colhead{Fe (solar)} 
& \colhead{log($n_{\rm e}t$/cm$^{-3}$\,s)} 
& \colhead{$L_{\rm X}$ (10$^{40}$ erg s$^{-1}$)}  
& \colhead{$\chi^2$ or C-value / d.o.f.}  
}
\startdata
52--68 (1) & 
{\tt TBabs}$\times(${\tt vpshock} + {\tt gaussian}$^d$) & 
44.5$^{+5.3}_{-4.8}$ & 
20$^c$ & 
0.4$^a$ & 
$>$11.95 &
67.3$^{+5.4}_{-5.1}$ &
88.9 / 79 \\
205--209 (1) & 
{\tt TBabs}$\times${\tt vpshock} & 
18.6$^{+9.4}_{-6.5}$ & 
20$^c$ & 
0.4$^a$ & 
12.3$^c$ &
56.8$^{+14.4}_{-12.4}$ & 
5.8 / 9 \\
383 (2) & 
{\tt TBabs}$\times${\tt vpshock} + {\tt TBabs}$\times${\tt vpshock} & 
0.93$^{+0.42}_{-0.36}$, 12.9$^{+2.4}_{-1.9}$ & 
20$^c$ & 
0.4$^c$ & 
$>$12.1 &
3.9$\pm$0.6, 57.2$\pm$3.5 & 
197.9 / 187 \\
620 (2) & 
{\tt TBabs}$\times${\tt vpshock} + {\tt TBabs}$\times${\tt vpshock} & 
$<$0.18, 5.0$^{+1.4}_{-1.0}$ & 
20$^c$ & 
0.4$^c$ & 
$>$12.4 &
4.8$\pm$0.5, 23.8$\pm$2.0 & 
52.8 / 55 \\
764 (3) & 
{\tt TBabs}$\times${\tt vpshock} & 
0.42$^{+0.16}_{-0.12}$ & 
20$^c$ & 
0.4$^a$ & 
$>$12.0 &
14.1$\pm$0.9 & 
81.4 / 42 \\
843--910 (3) & 
{\tt TBabs}$\times${\tt vpshock} & 
$<$0.12 & 
20$^c$ & 
0.4$^a$ & 
$>$12.3 &
8.0$^{+1.1}_{-1.0}$ & 
59.4 / 42 \\
1129 (3) & 
{\tt TBabs}$\times${\tt vpshock} & 
0.18$^{+0.04}_{-0.03}$ & 
20$^c$ & 
0.4$^a$ & 
$>$12.7 &
5.9$\pm$0.2 & 
231.3 / 166 \\
1340 (3) & 
{\tt TBabs}$\times${\tt vpshock} & 
0.53$^{+0.20}_{-0.18}$ & 
20$^c$ & 
0.4$^a$ & 
$>$12.7 &
5.9$\pm$0.2 & 
231.3 / 166 \\
1521--1545 (3) & 
{\tt TBabs}$\times${\tt vpshock} & 
$<$1.8 & 
20$^c$ & 
0.4$^a$ & 
12.3$^c$ &
2.6$^{+1.6}_{-1.2}$ & 
1.7 / 3 \\
\enddata
\tablecomments{$^d$The Gaussian centroid is obtained to be 6.34$\pm$0.05\,keV which corresponds to 6.41$\pm$0.05\,keV after correcting the redshift of the host galaxy.  Its width is set to zero.}
\end{deluxetable}

\end{document}